\documentclass[a4paper,11t]{extarticle}
\pdfoutput=1 

\usepackage{jheppub} 

\usepackage[OT1]{fontenc} 
\graphicspath{{Figures/}} 	     
\usepackage{booktabs} 
\usepackage{bbold} 
\usepackage{mathtools}
\usepackage{multirow}
\usepackage[version=4]{mhchem} 
\usepackage{physics}
\usepackage{pdflscape}
\usepackage{slashed}
\usepackage{xcolor}
\usepackage{tikz}
\usepackage{tikz-feynman}
\usepackage{makecell}

\usetikzlibrary{calc}

\def\vev{{\it vev}}
\def\vevs{{\it vevs}}

\usepackage{enumitem}

\usepackage{orcidlink}

\newcommand{\bea} {\begin{eqnarray}}
\newcommand{\eea} {\end{eqnarray}}

\newcommand{\be}{\begin{eqnarray}}
\newcommand{\ee}{\end{eqnarray}}

\newcommand{\beq}{\begin{equation}}
\newcommand{\eeq}{\end{equation}}

\newcommand{\Eq}[1]{Eq.~(\ref{#1})}

\def\vw{v_w}

\def\etmiss{\slashed{E}_T}

\title{Dark Matter and Electroweak Baryogenesis with Spontaneous \boldmath{$CP$} Violation in the Early Universe}

  \author[a]{Subhojit Roy\,\orcidlink{0000-0001-6434-5268}}
\affiliation[a]{HEP Division, Argonne National Laboratory, 9700 Cass Ave., Argonne, IL 60439, USA}
   \emailAdd{sroy@anl.gov}
  \abstract{Dark matter (DM) and the baryon asymmetry of the universe (BAU) are among the most compelling indications of physics beyond the Standard Model. We revisit the inelastic Higgs-portal complex singlet, a minimal framework in which a complex scalar splits into two nearly degenerate real states, with an off-diagonal Higgs-portal interaction that drives coannihilation to set the relic density, while the elastic DM-Higgs coupling can be tuned small enough to evade direct-detection limits. This setup naturally supports a strong first-order electroweak phase transition (SFOEWPT) and can account for the long-standing Galactic Center gamma-ray excess (GCE) via present-day DM annihilation into Higgs pairs. In this work, we show that the same framework, extended by a $Z_2$-symmetric dimension-6 $CP$-violating top Yukawa operator, can also generate the BAU via the electroweak baryogenesis (EWBG) mechanism. The cosmological history involves a two-step electroweak phase transition: first, the singlet fields acquire nonzero vacuum expectation values ($\it vevs$); then a strongly first-order transition occurs in which the Higgs develops its  nonzero  $\it vev$ while the singlet $\it vevs$ vanish. During this second step, both fields remain nonzero only within the advancing bubble wall, generating wall-localized $CP$ violation that biases sphaleron transitions and enables EWBG. After the phase transition, $CP$ and $Z_2$ symmetries are restored: the lightest singlet state becomes a stable DM candidate, while the vanishing singlet $\it vevs$ allow the model to naturally satisfy the stringent constraints on $CP$ violation. We delineate the SFOEWPT-favored parameter space, identifying the criteria for the two-step phase transition region that simultaneously yields the observed BAU and relic density, explains the GCE, and predicts gravitational wave spectra accessible to next-generation space-based detectors.}

\begin{document}
\maketitle
\flushbottom 
%
%
\section{Introduction}
The nature of dark matter (DM) and the origin of the baryon asymmetry of the universe (BAU) remain two of the most pressing open problems in modern physics. Observational evidence for DM is well established, from galaxy rotation curves~\cite{Rubin:1970zza} and gravitational lensing~\cite{Jee:2007nx} to anisotropies of the cosmic microwave background (CMB)~\cite{Planck:2015fie}, indicating a nonluminous component that accounts for about $26.4\%$ of the present cosmic energy budget~\cite{ParticleDataGroup:2024cfk}. At the same time, precision CMB analyses infer a baryon-to-entropy ratio
\begin{equation}
Y_B \equiv \frac{n_B - n_{\bar B}}{s} \simeq 8.65 \times 10^{-11},
\end{equation}
consistent across multiple datasets~\cite{Planck:2015fie,Planck:2018vyg,WMAP:2012nax,Steigman:2007xt}. Neither of these phenomena can be explained within the Standard Model (SM), necessitating physics beyond the SM (BSM) that can account for both the DM abundance and the BAU.

For the dynamical generation of the BAU, the three Sakharov conditions must be satisfied~\cite{Sakharov:1967dj}: (i) baryon number violation ($\Delta B\neq 0$), (ii) violation of $C$ and $CP$ symmetries, and (iii) a departure from thermal equilibrium. Several well-motivated frameworks realize these ingredients, including leptogenesis~\cite{Fukugita:1986hr,DAmbrosio:2003nfv,Pilaftsis:2003gt}, the Affleck--Dine mechanism~\cite{Affleck:1984fy,Dine:1995kz}, gravitational baryogenesis~\cite{Davoudiasl:2004gf}, and electroweak baryogenesis (EWBG)~\cite{Kuzmin:1985mm,Shaposhnikov:1986jp,Shaposhnikov:1987tw}. Among these, EWBG is particularly attractive: it requires new physics near the electroweak scale--the very regime under scrutiny at current and future collider experiments.

In the EWBG framework, a strong first-order electroweak phase transition (SFOEWPT) provides the requisite departure from equilibrium via expanding bubbles of the broken phase. $CP$-violating interactions in (or on) the bubble wall generate chiral charge densities that diffuse into the unbroken phase, where baryon-number-violating electroweak sphaleron processes convert them into a net baryon asymmetry~\cite{Klinkhamer:1984di,Gavela:1994dt,Huet:1994jb}. Once the bubbles expand and fill the universe, the sphaleron rate in the broken phase becomes exponentially suppressed, thereby freezing in the generated asymmetry. However, the SM itself cannot realize EWBG. Lattice and perturbative analyses show that, for the observed SM-like Higgs ($h$) mass $m_{h} \simeq 125~\mathrm{GeV}$~\cite{ATLAS:2012yve,CMS:2012qbp}, the electroweak phase transition (EWPT) is a crossover rather than first-order~\cite{Kajantie:1996mn,Csikor:1998eu,Aoki:1999fi}; the Higgs quartic coupling $\lambda_h$ suppresses the thermal cubic term, eliminating the barrier needed for an SFOEWPT~\cite{Bochkarev:1987wf,Kajantie:1995kf}. In addition, $CP$ violation from the Cabibbo–Kobayashi–Maskawa (CKM) phase is far too small to account for the observed asymmetry~\cite{Shaposhnikov:1987tw,Farrar:1993hn,Gavela:1993ts,Konstandin:2003dx,Kapusta:2006pm}. Consequently, achieving a strongly first-order transition and sufficient $CP$ violation for EWBG requires BSM physics. This mechanism has motivated extensive study of the EWPT and its realizations in BSM theories~\cite{Cohen:1993nk,Rubakov:1996vz,Trodden:1998ym,Riotto:1998bt,Cline:2006ts,Morrissey:2012db,White:2016nbo,Cline:2017jvp,Wagner:2023vqw,Liu:2023sey, Goncalves:2023svb,vandeVis:2025efm}. Despite its appeal, no conclusive evidence for new electroweak-scale excitations has yet emerged. 
Furthermore, any additional $CP$-violating interactions introduced at the electroweak scale are strongly constrained by precision measurements of electric dipole moments (EDM) of the electron~\cite{ACME:2018yjb}, neutron~\cite{Abel:2020pzs}, and heavy atoms such as mercury~\cite{Griffith:2009zz}. A viable way to reconcile these limits with successful EWBG is through spontaneous $CP$ violation that occurs only during the EWPT. In this picture, the temporary breaking of $CP$ in the early universe supplies the baryogenesis source while the symmetry is restored at zero temperature, naturally evading the present EDM constraints.

While EWBG highlights the need for new electroweak-scale physics, DM provides an independent window into such possibilities. Among the many candidates, weakly interacting massive particles (WIMPs) have long been attractive, motivated by the “WIMP miracle,” wherein weak-scale masses and couplings naturally yield the observed relic abundance via thermal freeze-out~\cite{Jungman:1995df, Griest:2000kj, Bertone:2004pz, Arcadi:2017kky}. However, null results from direct detection experiments~\cite{XENON:2018voc,XENON:2023cxc,LZ:2022lsv,LZCollaboration:2024lux} and collider searches~\cite{ATLAS:2021kxv,CMS:2021far}, together with strong bounds from indirect searches~\cite{Fermi-LAT:2015att}, have placed severe constraints on this paradigm. Intriguingly, the Galactic Center Excess (GCE) of gamma-ray observed by the Fermi Large Area Telescope may still be interpreted as a possible indirect signal of DM annihilation with a thermal cross section~\cite{Goodenough:2009gk,Hooper:2010mq,Abazajian:2012pn}, though astrophysical origins such as unresolved millisecond pulsars remain viable alternatives~\cite{Abazajian:2014fta,Lee:2015fea}.

One of the simplest WIMP realizations is the Higgs portal $Z_2$-symmetric real singlet scalar~\cite{Silveira:1985rk, McDonald:1993ex, Espinosa:1993bs, Burgess:2000yq, Barger:2007im, Barger:2008jx, Ashoorioon:2009nf, Cline:2012hg, Cline:2013gha, Cheung:2013dca, Jiang:2015cwa, Chala:2016ykx, Arcadi:2017kky, Kurup:2017dzf, GAMBIT:2017gge, Grzadkowski:2018nbc, Funakubo:2025utb}, in which the DM couples quadratically to the Higgs boson and freezes out with the observed relic density. Aside from the Higgs-resonant region at $m_\phi \simeq m_h/2$, where the relic abundance can be reproduced with relatively small Higgs–portal coupling, this scenario is excluded across most of its parameter space by the combined impact of direct detection~\cite{Cline_2013,Casas:2017jjg,DiazSaez:2024nrq}, indirect detection~\cite{DeLaTorreLuque:2023fyg}, and collider searches~\cite{Djouadi:2011aa,Arcadi:2019lka,Krnjaic:2015mbs}. Similarly, the $Z_2$-symmetric complex singlet with only elastic portal interactions with the SM Higgs is also excluded, apart from the Higgs pole region. Both of these extensions have been studied in detail~\cite{McDonald:1993ey,McDonald:1995hp,Barger:2008jx,Profumo:2007wc,Gonderinger:2012rd,Jiang:2015cwa,Chao:2017oux,Chiang:2017nmu,Cline:2012hg,Grzadkowski:2018nbc} in connection with DM phenomenology as well as EWBG. The presence of additional scalar degrees of freedom can alter the Higgs potential at finite temperature through thermal corrections, effectively generating cubic terms that favor an SFOEWPT. However, realizing such a transition typically requires sizeable Higgs–portal couplings and an electroweak-scale singlet, which are now strongly constrained by direct-detection limits. Consequently, these minimal singlet models cannot simultaneously provide a viable DM candidate (except near the Higgs pole) and realize an SFOEWPT, and thus fail to accommodate EWBG.

Motivated by these considerations, recent work indicates that electroweak–scale DM can be reconciled with an SFOEWPT in a minimally extended Higgs–portal framework with a complex scalar~\cite{Hooper:2025fda}. An approximate global $U(1)$ carried by $\phi$ is explicitly broken by mass terms, splitting the field into two non-degenerate real states $\phi_{1,2}$. 
In the mass basis, the leading Higgs–portal interaction can take the off-diagonal leading form,
\begin{equation}
\mathcal{L}\supset g \phi_1\phi_2 H^\dagger H \,,
\end{equation}
such that direct detection is dominated by inelastic scattering, while the elastic rate is suppressed by keeping the elastic coupling tiny. Here, $H$ denotes the SM Higgs doublet. The correct relic abundance can still be obtained through the well-known co-annihilation mechanism.
It has been further shown in our previous work that the same structure can accommodate an interpretation of the GCE excess via annihilation into Higgs pairs ($\phi_1\phi_1\to hh$) without requiring large elastic couplings and remains compatible with present bounds. In this scenario, the heavier singlet state, $\phi_2$, can sustain a relatively large portal coupling to the Higgs, since it contributes to the spin–independent DM–nucleon cross section only at one loop. Such a coupling can significantly reshape the finite–temperature Higgs potential and favor an SFOEWPT, which can source stochastic gravitational wave (GW) background at the early universe. 

In the present work, we build on that foundation by studying EWBG within this framework. Among the various patterns of phase transition, the most relevant for EWBG is the two-step sequence in which, at higher temperatures, the minimum first shifts along the singlet directions with $\langle\phi_{1,2}\rangle\neq 0$, and at lower temperatures an SFOEWPT occurs in which the Higgs acquires a nonzero vacuum expectation value ($\vev$) while the singlet expectations return to zero, thereby restoring the imposed $Z_2$. During the intermediate stage, the singlet background develops a complex phase across the bubble wall, which can source $CP$ violation. To realize this explicitly, we extend the model by including the dimension–6 operator
\begin{equation}
\mathcal{O}_{(6)} \,\supset\, y_t \, \overline{Q} \, \widetilde{H} \!\left( 1 +  c\, \frac{\phi^2}{\Lambda^2} \right) t_R + \text{h.c.},
\end{equation}
with 
$\widetilde H=i\sigma_2 H^*$ and $\Lambda$ a cutoff scale~\cite{McDonald:1993ey,Chao:2017oux,Comelli:1993ne}.~Here, $Q$ and $t_R$ are the third-generation left-handed quark doublet and the right-handed top quark, respectively. The coefficient `$c$' is a dimensionless Wilson coefficient.~This operator provides the needed new source of $CP$ violation during the phase transition: the phase of $\phi$ across the bubble wall generates chiral charge densities that diffuse into the unbroken phase and are partially converted into baryon number by electroweak sphalerons.~After the transition completes, the singlet expectation value vanishes, the $Z_2$ and $CP$symmetries are restored, the lighter real component is stabilized as the DM candidate, and stringent EDM constraints are naturally avoided due to the absence~of~zero–temperature~$CP$~violation~\cite{ACME:2013pal,ParticleDataGroup:2024cfk,ACME:2018yjb,Abel:2020pzs,Griffith:2009zz}.

Beyond baryogenesis, an FOEWPT in such models can also act as a powerful source of stochastic GW in the early universe. These arise from bubble collisions, sound waves, and magnetohydrodynamic (MHD) turbulence, and the predicted spectra may lie within the reach of upcoming space–based interferometers such as LISA, Taiji, BBO, and UDECIGO, offering a complementary observational window into electroweak–scale physics~\cite{Caprini:2015zlo,Cai:2017cbj,Caprini:2018mtu,Romano:2016dpx,Christensen:2018iqi,Athron:2023xlk}.

In this work we study this inelastic Higgs–portal model with transient $CP$ violation and assess its ability to account simultaneously for the DM relic abundance, EWBG, the observed GCE, and associated GW signatures. While the DM phenomenology of this framework has been analyzed in detail in Refs.~\cite{Ghorbani:2014gka,Casas:2017jjg, Guo:2021vpb, DiazSaez:2024nrq,Hooper:2025fda, Guo:2025qes, Goncalves:2025snm}, here we revisit it to highlight the features most relevant for EWBG and to clarify their interplay with DM physics. We first analyze the scalar potential and its finite–temperature corrections that enable a first-order transition, and then examine the DM relic density and direct–detection limits.
Building on this, we identify the parameter space consistent with an SFOEWPT and a two–step phase–transition scenario, determine the resulting GW spectrum, and assess the generation of the baryon asymmetry through dynamical $CP$ violation localized on the expanding bubble walls. Our results show that, despite stringent experimental constraints, this minimal scalar extension can simultaneously accommodate a viable cold DM candidate, generate the observed BAU, provide a GCE-compatible gamma–ray signal, and produce GW spectra testable at next-generation detectors~\footnote{While completing this work, Ref.~\cite{Roux:2025wem} appeared, which studies a different DM model addressing both the GCE and EWBG.}.

The paper is organized as follows. Sec.~\ref{theoretical} presents the inelastic Higgs–portal DM framework, constructs the finite–temperature effective potential, reviews the general requirements for EWBG, and summarizes GW production from a first-order phase transition (FOPT). DM phenomenology is discussed in Sec.~\ref{relicdm}. In Sec.~\ref{foptregion} we explore the parameter space that favors an FOEWPT including two-step phase transition scenario in the early universe. The generation of the BAU via spontaneous $CP$ violation in this model is analyzed in Sec.~\ref{BAU}. Finally, we conclude in Sec.~\ref{Conclusions}.
\section{Theoretical Framework}
\label{theoretical}
\subsection{The Inelastic Higgs-Portal Dark Matter Model}
We begin by recalling the inelastic Higgs-portal DM model, previously analyzed in Ref.~\cite{Hooper:2025fda}, in order to make the present discussion self-contained. 
The model extends the SM with a complex scalar singlet $\phi$ that couples only to the Higgs doublet $H$. 
The tree-level scalar potential involving $H$ and $\phi$ is given by~\cite{Hooper:2025fda}:
\begin{align}
\label{Vtreetot}
V(H,\phi) &= - \mu_h^2 H^\dagger H + \lambda_h (H^\dagger H)^2 + m_0^2 |\phi|^2 + \frac{1}{2}\left( \rho_0^2 \phi^2 + \rho{_0^{2}}{^{*}} \phi^{*2} \right) \nonumber \\
&~~~ + \left( \kappa |\phi|^2 + \frac{1}{2} \left( \eta \phi^2 + \eta^* \phi^{*2} \right) \right) H^\dagger H,
\end{align}
where $\rho_0^2$ and $\eta$ can, in general, be complex parameters. 
The complex scalar $\phi$ approximately respects a global $U(1)$ symmetry, explicitly broken by the $\phi^2$ terms, reducing the symmetry to a discrete $Z_2$ under which $\phi \rightarrow -\phi$.
After electroweak symmetry breaking (EWSB), the Higgs doublet can be expanded around the vacuum as
\[
H = \left(G^{+}, \frac{1}{\sqrt{2}}(v_h + h + i G^0)\right)^T,
\]
where $v_h = 246$~GeV is the vacuum expectation value ($\vev$) of the neutral $CP$-even Higgs component at zero temperature.
The mass eigenstates of the scalars can be obtained through the diagonalization of the mass-matrix. The eigenvalues of the mass eigenstates
of the singlet sector, denoted as $\phi_1$ and $\phi_2$, are given by,
\be
m^2_{\phi_1, \phi_2} = m^2 \mp |\rho^2|.
\ee
In the limit of a small mass splitting, we have
\be
 \Delta m = m_{\phi_2} - m_{\phi_1} \approx \frac{|\rho^2|}{ m},
\ee 
where,
\be
m^2 \equiv m_0^2 +\frac{ \kappa v_h^2 }{2},~~
\rho^2 \equiv \rho_0^2 + \frac{\eta v_h^2}{2}\, .
\ee
Here, $\rho^2$ is a complex parameter since both $\rho_0^2$ and $\eta$  are complex. These parameters can be decomposed into their real and imaginary parts as follows:
\be
\rho_0^2 = \rho_{0_R}^2 + i \rho_{0_I}^2 \, , ~~~~~ \quad \eta = \eta_R + i \eta_I \, , ~~~~~ \rho^2 = \rho_R^2 + i \rho_I^2 \, \,. ~~~~~~
\ee
The complex field $\phi$ can be decomposed in terms of the mass eigenstates as:
\begin{equation}
\phi = \alpha \phi_1 + \beta \phi_2, \qquad
\phi^* = \alpha^* \phi_1 + \beta^* \phi_2,
\end{equation}
where $\alpha = e^{-i\theta}/\sqrt{2}$ and $\beta = i e^{-i\theta}/\sqrt{2}$. The mixing angle $\theta$ diagonalizes the singlet mass matrix via the rotation:
\begin{equation}
\label{eqn:mneut}
{\cal U} = \begin{pmatrix}
\cos\theta & -\sin\theta \\
\sin\theta & \cos\theta
\end{pmatrix},
\end{equation}
with 
\begin{equation}
\cos 2\theta = \frac{\rho_R^2}{|\rho^2|} \, , \qquad \sin 2\theta = \frac{\rho_I^2}{|\rho^2|}.
\end{equation}
After EWSB, the Higgs portal interactions in the mass basis take the form:
\begin{equation}
\label{eq:Lh-mass}
{\cal V}_{\phi h} = \left(f_1 \phi_1^2 + f_2 \phi_2^2 + g \phi_1 \phi_2\right)\left(-v_h h + \frac{1}{2} h^2\right),
\end{equation}
where the couplings are given by:
\begin{align}
f_1 &= \frac{\kappa}{2} + \frac{1}{2}(\eta \alpha^2 + \eta^* \alpha^{*2}) = \frac{\kappa}{2} + \frac{1}{2} \eta_R \cos 2\theta + \frac{1}{2} \eta_I \sin 2\theta \, \, , \label{f1rel}\\
f_2 &= \frac{\kappa}{2} + \frac{1}{2}(\eta \beta^2 + \eta^* \beta^{*2}) = \frac{\kappa}{2} - \frac{1}{2} \eta_R \cos 2\theta - \frac{1}{2} \eta_I \sin 2\theta \, \, , \label{f2rel} \\
g   &= \eta \alpha \beta + \eta^* \alpha^* \beta^* = \eta_R \sin 2\theta - \eta_I \cos 2\theta \, \,. \label{grel}
\end{align}
Here, $f_1$ and $f_2$ control the diagonal (elastic) interactions, while $g$ governs the off-diagonal (inelastic) Higgs portal coupling between $\phi_1$ and $\phi_2$.
This setup differs significantly from the conventional real singlet scalar DM~\cite{Silveira:1985rk, McDonald:1993ex, Espinosa:1993bs, Burgess:2000yq, Barger:2007im, Barger:2008jx, Ashoorioon:2009nf, Cline:2012hg, Cline:2013gha, Cheung:2013dca, Jiang:2015cwa, Arcadi:2017kky, Kurup:2017dzf, GAMBIT:2017gge, Grzadkowski:2018nbc, Funakubo:2025utb}, especially in the regime where $|f_1| \ll |f_2|, |g|$. Notably, this model allows for a viable DM candidate even in the limit $f_1 \rightarrow 0$.

Without loss of generality, to study DM and EWPT phenomenology, one can consider a particular value of the diagonalizing angle $\theta$. 
In this work, we focus on the case $\sin 2\theta = 0$, which implies:
$\rho_{0_I}^2 = -\frac{v_h^2}{2} \eta_I$.
In this limit, the tree-level scalar potential at zero temperature becomes:
\begin{align}
\label{Vtree}
V_0(h, \phi_1, \phi_2) = & -\frac{1}{2} \mu_h^2 h^2 + \frac{1}{4} \lambda_h h^4 + \frac{1}{2} \left(m_0^2 + \frac{\kappa}{2} h^2 \right)(\phi_1^2 + \phi_2^2)  + \frac{1}{2} \left( \rho_{0_R}^2 + \frac{\eta_R}{2} h^2 \right)(\phi_1^2 - \phi_2^2) \nonumber \\
& - \frac{\eta_I}{2} (h^2 - v_h^2) \phi_1 \phi_2  + \frac{1}{4} \left( \lambda_1 \phi_1^4 + \lambda_2 \phi_2^4 + \lambda_{12} \phi_1^2 \phi_2^2 \right) .
\end{align}
We add quartic self-interactions for the singlet fields $\phi_1$ and $\phi_2$, parametrized by $\lambda_1$, $\lambda_2$, and $\lambda_{12}$. In the special case $\lambda_1 = \lambda_2 = \lambda_{12}$, the scalar sector retains an approximate $U(1)$ symmetry. More generally, allowing these couplings to differ explicitly breaks this symmetry.
In the limits of $f_1 \sim 0$ and $\sin2\theta=0$, the coupling relations given in Eq.~\eqref{f1rel} to Eq.~\eqref{grel} and the mass relation are reduced to, 
\begin{align}
\eta_I = - g, \, \, \, \quad \kappa = f_2, \, \, \, \eta_R = -\kappa, \, \, \,  
m_0^2 = \frac{1}{2}(m_{\phi_1}^2 + m_{\phi_2}^2 - \kappa v_h^2) , 
\, \, \,  \rho_{0_R}^2 = \frac{1}{2}(m_{\phi_1}^2 - m_{\phi_2}^2 
+ \kappa v_h^2
).\label{rho0Rsq}
\end{align}
Since the potential respects a $Z_2$ symmetry, the mass eigenstates $\phi_1$ and $\phi_2$ transform identically under this symmetry. Consequently, the lighter state $\phi_1$ is stable and serves as the DM candidate. The vacuum structure at zero temperature is:
\begin{equation}
\langle h \rangle = v_h, \qquad \langle \phi_1 \rangle = 0, \qquad \langle \phi_2 \rangle = 0.
\end{equation}
From a phenomenological standpoint, the relevant set of independent parameters in the scalar potential~\eqref{Vtree} can be taken as:
\begin{equation}
\label{free-para}
\left\{ \, \, m_{\phi_1}, \,  \Delta m, \, f_1, \, f_2, \, g, \,\lambda_1, \, \lambda_2, \, \lambda_{12} \, \,  \right\}.
\end{equation}

\subsection{Finite Temperature Corrections to the Effective Potential}

Understanding the dynamics of phase transitions in the early universe requires studying the behavior of the scalar potential at nonzero temperatures. The starting point is the classical scalar potential at zero temperature, $V_0(h, \phi_1, \phi_2)$ (see Eq.~\eqref{Vtree}), which encodes the tree-level interactions of the scalar fields relevant for EWSB.

\subsubsection{Quantum Corrections at Zero Temperature}

Quantum effects modify the classical potential via loop corrections. At one-loop order, these corrections are encapsulated by the Coleman-Weinberg (CW) potential~\cite{Coleman:1973jx}. The zero-temperature one-loop effective potential can be expressed as
\begin{equation}
\label{eq:VCW_rewrite}
V_{\rm CW}(h, \phi_1, \phi_2) = \sum_i (-1)^{2 s_i} \frac{n_i}{64 \pi^2} m_i^4(h, \phi_1, \phi_2) \left[ \ln\left(\frac{m_i^2(h, \phi_1, \phi_2)}{Q^2}\right) - C_i \right],
\end{equation}
where the index $i$ runs over all particle species in the theory. Here, $m_i(h, \phi_1, \phi_2)$ are the field-dependent masses, $s_i$ and $n_i$ denote the spin and number of degrees of freedom (d.o.f.) of each species, respectively. The scale $Q$ is the renormalization scale, which we set equal to the electroweak vacuum expectation value $v$. The constants $C_i$ depend on the renormalization scheme; in the $\overline{\text{MS}}$ on-shell scheme used here, $C_i=5/6$ for transverse gauge bosons and $C_i=3/2$ for longitudinal modes, scalars, and fermions.
For clarity, the relevant degrees of freedom included in the sum are:
\begin{equation}
\begin{aligned}
&n_h = n_{\phi_1} = n_{\phi_2} = n_{G^0} = 1, \quad n_{G^\pm} = 2, \quad  n_Z = 3, \quad n_{W^\pm} = 6, \quad n_t = 12\, .
\end{aligned}
\end{equation}
In this work, we choose to work in the Landau gauge, which simplifies the treatment by decoupling ghost fields.

One-loop corrections shift both the location of the electroweak vacuum and the scalar masses and mixings. To preserve the physical mass spectrum and vacuum expectation values at one loop, counterterms $V_{\rm CT}$ are introduced, parametrized by
\begin{equation}
\begin{aligned}
V_{\rm CT} = &- \frac{1}{2} \delta \mu_h^2 h^2 + \frac{1}{4} \delta \lambda_h h^4 + \delta m_{12}^2 h \phi_1 + \delta m_{13}^2 h \phi_2  + \delta m_{23}^2 \phi_1 \phi_2 + \frac{1}{4} \delta \lambda_{12} h^2 \phi_1^2 + \frac{1}{4} \delta \lambda_{13} h^2 \phi_2^2.
\end{aligned}
\end{equation}
The coefficients of these counterterms are fixed by imposing on-shell renormalization conditions at zero temperature, ensuring that the first and second derivatives of the total one-loop potential (including counterterms) vanish at the electroweak vacuum $\{h, \phi_1, \phi_2\} = \{v, 0, 0\}$:
\begin{equation}
\label{eq:renorm_conditions}
\left. \frac{\partial (V_{\rm CW} + V_{\rm CT})}{\partial \phi_i} \right|_{\rm vacuum} = 0, \quad
\left. \frac{\partial^2 (V_{\rm CW} + V_{\rm CT})}{\partial \phi_i \partial \phi_j} \right|_{\rm vacuum} = 0,
\end{equation}
where $\phi_i, \phi_j \in \{h, \phi_1, \phi_2\}$.
The relations of various coefficients of $V_{\rm CT}$  are given by,
\be
\delta\mu_h^2  &=& \frac{3}{2 v} DV[1] - \frac{1}{2}DV[1,1] ~~,~~
\delta \lambda_{h} =\frac{1}{2 v^3} DV[1] - \frac{1}{2 v_h^2}DV[1,1], 
~~~~~~~~\\
\delta m_{12}^2 &=& - \frac{1}{v} DV[2] ~~,~~
\delta m_{23}^2 = -  DV[2,3] ~~,~~
\delta m_{13}^2 = - \frac{1}{v} DV[3] ~~~~
\\
\delta \lambda_{12} &=& - \frac{2}{v_h^2} DV[2,2]~~~,~~~
\delta \lambda_{13} = - \frac{2}{v_h^2} DV[3,3]  ,
\ee
where we have defined
\be
DV[i] \equiv \frac{\partial V_\text{CW}}{\partial s_{_i}} ~~~,~~~ DV[i,j] \equiv \frac{\partial^2 V_\text{CW}}{\partial s_{_i}  \partial s_{_j}},
\ee
for $s_{_i}, s_{_j} = \{h, \phi_1, \phi_2\}$. All derivatives are taken at the true electroweak minima, i.e., $h = v$, $\phi_1 = 0$ and $\phi_2 = 0$.
A subtlety arises due to the massless Goldstone bosons at the vacuum in Landau gauge, causing infrared divergences in the second derivatives of $V_{\rm CW}$. To regulate these, we introduce a small IR regulator mass $\mu_{\rm IR}^2$ by shifting $m_G^2 \to m_G^2 + \mu_{\rm IR}^2$, with $\mu_{\rm IR} \approx 1\, \text{GeV}$, following established methods~\cite{Baum:2020vfl, Chatterjee:2022pxf, Hooper:2025fda, Bittar:2025lcr, Roy:2022gop, Ghosh:2022fzp}.

\subsubsection{Thermal Corrections}
\label{thermalcorr}
At finite temperature $T$, the effective potential receives additional contributions from the thermal bath. The one-loop thermal corrections~\cite{Dolan:1973qd, Weinberg:1974hy} take the form
\begin{equation}
\label{eq:Vthermal}
V_{\rm th}(h, \phi_1, \phi_2, T) = \frac{T^4}{2 \pi^2} \sum_i n_i J_{B,F} \left( \frac{m_i^2(h, \phi_1, \phi_2)}{T^2} \right),
\end{equation}
where the thermal functions $J_B$ and $J_F$ correspond to bosons and fermions, respectively:
\begin{equation}
J_{B,F}(y^2) = \int_0^\infty dx\, x^2 \ln\left[1 \mp \exp\left(-\sqrt{x^2 + y^2}\right)\right],
\end{equation}
with the upper sign for bosons and lower for fermions.
In the low-temperature limit, the thermal functions are exponentially suppressed. A convenient asymptotic form is
\begin{align}
    &\label{eq:JLTapprox}
    J_{B,F}(y^2) \big|_{LT} \approx - \left(\frac{\pi}{2}\right)^{1/2} y^{3/2} e^{-y} \left( 1+ \frac{15}{8} y^{-1}\right)
\end{align}
Thus, heavy modes with $m_i^2\gg T^2$ contribute only via Boltzmann tails and effectively decouple from the finite-temperature potential.

In the high-temperature regime ($m_i^2 \ll T^2$), the thermal functions admit expansions:
\begin{equation}
\begin{aligned}
J_B(y^2) &\approx -\frac{\pi^4}{45} + \frac{\pi^2}{12} y^2 - \frac{\pi}{6} y^3 - \frac{1}{32} y^4 \ln\left(\frac{y^2}{a_B}\right), \\
J_F(y^2) &\approx \frac{7 \pi^4}{360} - \frac{\pi^2}{24} y^2 - \frac{1}{32} y^4 \ln\left(\frac{y^2}{a_F}\right),
\end{aligned}
\end{equation}
where 
\begin{equation}
a_B = 16 \pi^2 \exp\left(\frac{3}{2} - 2 \gamma_E\right), \quad a_F = \frac{a_B}{16},
\end{equation}
and $\gamma_E \approx 0.577$ is the Euler-Mascheroni constant.

Of particular importance is the cubic term in $J_B$, which generates a term $\propto - T m^3$ in the effective potential. This term can induce an energy barrier between degenerate minima, enabling an FOPT. However, it also leads to infrared divergences that signal the breakdown of naive perturbation theory at high temperature.

\subsubsection{Daisy Resummation}

To cure the infrared problems associated with the bosonic zero Matsubara modes, one must resum a class of higher-loop diagrams known as the daisy or ring diagrams~\cite{Gross:1980br, Parwani:1991gq, Arnold:1992rz}. In this work, we implement the Parwani resummation scheme~\cite{Parwani:1991gq}, wherein the bosonic field-dependent masses $m_i^2$ are replaced by thermally corrected masses
\begin{equation}
M_i^2(h, \phi_1, \phi_2, T) = m_i^2(h, \phi_1, \phi_2) + \Pi_i(T),
\end{equation}
with $\Pi_i(T) = c_i T^2$ representing the thermal mass corrections, whose coefficients $c_i$ depend on the particle content and interactions.
The full one-loop finite-temperature effective potential, including counterterms and daisy resummation~\footnote{It is well-known that the dynamics of a phase transition can be significantly affected by the choice of thermal resummation scheme, such as the Parwani~\cite{Parwani:1991gq}, Arnold-Espinosa~\cite{Arnold:1992rz}, full dressing, partial dressing or tadpole resummation~\cite{Boyd:1993tz} approaches. Recently, a detailed comparison of these resummation schemes has been performed for the first time in the context of a realistic BSM scenario, such as the Two-Higgs-Doublet Model~\cite{Bittar:2025lcr}. In this work, we adopt the Parwani scheme for our analysis, while leaving the exploration of alternative schemes for future study. For further details, see references within~\cite{Bittar:2025lcr}.},
 is then
\begin{equation}
\label{eq:Veff_full}
V_{\rm eff}(h, \phi_1, \phi_2, T) = V_0(h, \phi_1, \phi_2) + V_{\rm CW}(M_i^2) + V_{\rm CT} + V_{\rm th}(M_i^2, T).
\end{equation}

\subsubsection{Scalar and Gauge Boson Masses at Finite Temperature}

To accurately analyze the behavior of the effective potential at finite temperature, it is crucial to incorporate the thermal corrections to the field-dependent masses of the scalar and gauge bosons. These thermal mass corrections directly influence the dynamics and nature of the EWPT.


The $CP$-even scalar fields $\{h, \phi_1, \phi_2\}$ contribute to a symmetric $3 \times 3$ field-dependent mass-squared matrix, denoted as $m_H^2(T)$. The diagonal elements are given by:
\begin{align}
\label{mH11sq}
m_{H_{11}}^2 &= -\mu_h^2 + 3 \lambda_h h^2 + \frac{\eta_I}{2} \phi_1 \phi_2 + \frac{\kappa}{2} (\phi_1^2 + \phi_2^2) + \frac{\eta_R}{2}(\phi_1^2 - \phi_2^2), \\
\label{mH22sq}
m_{H_{22}}^2 &= m_0^2 + \frac{\kappa + \eta_R}{2} h^2 + \rho_{0_R}^2 + \frac{\lambda_{12}}{2} \phi_2^2 + 3 \lambda_1 \phi_1^2, \\
\label{mH33sq}
m_{H_{33}}^2 &= m_0^2 + \frac{\kappa - \eta_R}{2} h^2 - \rho_{0_R}^2 + \frac{\lambda_{12}}{2} \phi_1^2 + 3 \lambda_2 \phi_2^2.
\end{align}
The off-diagonal terms describing mixing between the scalar fields are:
\begin{align}
m_{H_{12}}^2 &= m_{H_{21}}^2 = \kappa h \phi_1 + (\eta_R - \eta_I) h \phi_2, \\
m_{H_{13}}^2 &= m_{H_{31}}^2 = \kappa h \phi_2 - \eta_R h \phi_1 - \eta_I h \phi_1, \\
m_{H_{23}}^2 &= m_{H_{32}}^2 = - \rho_{0_I}^2 - \frac{1}{2} \eta_I h^2 + \lambda_{12} \phi_1 \phi_2.
\label{mh23}
\end{align}
%
%
The tree-level field-dependent masses for the Goldstone bosons, electroweak gauge bosons, and the top quark are:
\begin{subequations}
\label{eq:gauge-top-masses}
\begin{align}
\label{mg0gpmsq}
m_{G^0,G^{\pm}}^2 &= - \mu_h^2 + \lambda_h h^2, \\
m_{W^{\pm}}^2 &= \frac{1}{4} g_2^2 h^2, \\
m_Z^2 &= \frac{1}{4} (g_1^2 + g_2^2) h^2, \\
m_t^2 &= \frac{1}{2} y_t^2 h^2.
\end{align}
\end{subequations}
%
%
At high temperatures, thermal corrections from plasma effects contribute significantly to the mass spectrum, especially for bosonic zero modes. These corrections are captured via daisy resummation and modify the effective mass-squared matrix as:
\begin{equation}
M^2_{\text{eff}}(T) = m^2 + \Pi(T^2),
\end{equation}
where $\Pi(T^2)$ represents the thermal self-energy correction. These corrections are parametrized as $\Pi_{kl}(T^2) = c_{kl} T^2$, with $c_{kl}$ referred to as daisy coefficients. They can be obtained from the high-temperature limit of the thermal potential:
\begin{equation}
c_{kl} = \left. \frac{1}{T^2} \frac{\partial^2 V_{\text{th}}}{\partial \phi_k \partial \phi_l} \right|_{T^2 \gg m^2}.
\end{equation}
The daisy coefficients for the $CP$-even scalar fields are given by:
\begin{subequations}
\label{eq:daisy-coefficients}
\begin{align}
\label{c11}
c_{{11}} &= \frac{1}{16} (3 g_2^2 + g_1^2) + \frac{1}{4} y_t^2 + \frac{1}{48} (24 \lambda_h + 4 \kappa - \eta_I), \\
\label{c22}
c_{{22}} &= \frac{1}{24} (8 \kappa + 4 \eta_R + 6 \lambda_1 + \lambda_{12}), \\
\label{c33}
c_{{33}} &= \frac{1}{24} (8 \kappa - 4 \eta_R + 6 \lambda_2 + \lambda_{12}).
\end{align}
\end{subequations}
The temperature-corrected mass-squared values for the scalars are obtained by adding $c_{{ii}} T^2$ to $m_{H_{ii}}^2$, defined from \Eq{mH11sq}--(\ref{mH33sq}). A similar correction applies to the Goldstone boson masses, defined in \Eq{mg0gpmsq}, using $c_{{11}} T^2$.

The longitudinal components of the electroweak gauge bosons acquire additional temperature-dependent contributions due to their interactions with the thermal bath. For the $W_L^{\pm}$ bosons, the thermally improved mass is:
\begin{equation}
M_{W_L^{\pm}}^2 = \frac{1}{4} g_2^2 h^2 + \frac{11}{6} g_2^2 T^2.
\end{equation}
The longitudinal components of the Z-boson and the photon ($\gamma$) fields also receive thermal corrections. Their combined mass matrix in the longitudinal sector is:
\begin{equation}
\frac{1}{4} h^2
\begin{pmatrix}
g_2^2 & -g_2 g_1 \\
-g_2 g_1 & g_1^2
\end{pmatrix}
+
\begin{pmatrix}
\frac{11}{6} g_2^2 T^2 & 0 \\
0 & \frac{11}{6} g_1^2 T^2
\end{pmatrix}.
\end{equation}
Diagonalizing this matrix yields the thermally corrected mass-squared eigenvalues for the longitudinal modes of the 
Z-boson and photon, given by:
\begin{align}
M^2_{Z_L,\gamma_L} &=
\frac{1}{2} \left[
\frac{1}{4} (g_1^2 + g_2^2) h^2 + \frac{11}{6} (g_1^2 + g_2^2) T^2
\pm \delta
\right].
\end{align}
where the splitting term $\delta$ is defined as: $\delta = \sqrt{
\left( \frac{1}{4} (g_2^2 - g_1^2) h^2 + \frac{11}{6} (g_2^2 - g_1^2) T^2 \right)^2
+ g_1^2 g_2^2 h^4 }$.
Tracking the evolution of the global minimum of $V_{\rm eff}$ with temperature allows one to study the nature of the EWPT. In particular, a strongly FOPT, as dictated by the shape of the thermal potential, has profound implications for both the generation of the BAU and the possible production of a stochastic background of GW. In the following sections, we explore how such a phase transition can serve as the necessary out-of-equilibrium condition for successful EWBG, and how the associated dynamics may leave observable imprints in the form of gravitational radiation detectable by future experiments.
\subsection{Generalities of Electroweak Baryogenesis}
\label{subsec:bg}
EWBG is an appealing mechanism for generating the BAU within a testable energy regime, centered around the electroweak scale. As with any viable baryogenesis scenario, it must satisfy the three Sakharov conditions: (i) baryon number violation, (ii) violation of $C$ and $CP$ symmetries, and (iii) departure from thermal equilibrium~\cite{Sakharov:1967dj}. In the EWBG framework, these conditions can naturally be realized in the context of an FOEWPT, which can occur during the spontaneous breaking of the electroweak symmetry as the universe cools.

At high temperatures in the early universe, the Higgs $\vev$ vanishes and the electroweak symmetry remains unbroken. As the temperature drops below a critical value, $T_c$, the shape of the finite-temperature effective potential $V_{\rm eff}(T)$ evolves due to thermal and radiative corrections~\cite{Dolan:1973qd,Weinberg:1974hy,Coleman:1973jx,Kapusta:2006pm}. An FOEWPT is characterized by the appearance of two distinct minima of the potential: one at the origin (symmetric phase) and another at nonzero field values (broken phase), separated by a potential barrier. When these two minima become degenerate in energy at $T=T_c$, bubble nucleation can begin.

As the temperature further decreases to a nucleation temperature $T_n < T_c$, bubbles of the true vacuum (broken phase) begin to form within the surrounding false vacuum (symmetric phase). These bubbles expand, collide, and eventually complete the phase transition. The nucleation rate per unit volume per unit time is given approximately by $\Gamma_B(T) \sim T^4 \exp[-S_3(T)/T]$, where $S_3(T)$ is the so-called three-dimensional Euclidean action evaluated along the bounce solution~\cite{Langer:1969bc,Coleman:1977py,Affleck:1980ac}. A successful transition requires that $\Gamma_B \sim H^4$, which is typically achieved when $S_3(T)/T \lesssim 140$~\cite{Linde:1981zj,Mazumdar:2018dfl}. We employ \texttt{CosmoTransitions}~\cite{Wainwright:2011kj} to compute the bounce profile and determine $T_n$.

In the presence of an FOEWPT, the three Sakharov conditions can be realized as follows:
\begin{itemize}
\item \textbf{Baryon number violation:} The SM already allows for baryon number violation through nonperturbative processes associated with the $SU(2)_L$ gauge anomaly, known as sphaleron transitions~\cite{tHooft:1976rip,Klinkhamer:1984di,Manton:1983nd}. These transitions connect vacua with different Chern-Simons numbers and induce $\Delta B \neq 0$ processes. In the symmetric phase, sphaleron rates are unsuppressed ($\Gamma_{\text{sph}} \propto T^4$)~\cite{Arnold:1987mh,Khlebnikov:1988sr}, while in the broken phase they are exponentially suppressed due to the large energy barrier set by the sphaleron energy $E_{\text{sph}}(T)$.
\item \textbf{$C$ and $CP$ violation:} Since sphaleron processes violate baryon number but conserve $B-L$, they affect baryons and antibaryons symmetrically in the absence of any charge or $CP$ asymmetry. Therefore, a successful realization of baryogenesis demands a source of $CP$ violation that biases these transitions to favor baryon over antibaryon production. During an FOEWPT, $CP$-violating interactions between fermions and the advancing bubble wall can generate chiral charge asymmetries in the symmetric phase ahead of the wall~\cite{Farrar:1993sp,Farrar:1993hn}. These asymmetries diffuse into the plasma and act as source terms for sphalerons, enabling net baryon production.

However, the $CP$ violation present in the SM, originating from the CKM matrix, is far too feeble to account for the observed BAU~\cite{Gavela:1994dt,Gavela:1994ds}. This inadequacy motivates the inclusion of additional sources of $CP$ violation, typically introduced in extensions of the SM.

In this work, we dynamically generate $CP$ violation by augmenting the theory with a dimension-6, $Z_2$-symmetric, $CP$-violating operator in the top-quark Yukawa sector:
\begin{equation}
    \mathcal{O}_{(6)} \supset y_t \, \overline{Q} \, \widetilde{H} \left( 1 +  c\, \frac{\phi^2}{\Lambda^2} \right) t_R + \text{h.c.},
    \label{dim-6cpviolation}
\end{equation}
where $y_t$ is the top Yukawa coupling, $c$ is a dimensionless parameter controlling the strength of $CP$ violation, and $\Lambda$ is the scale of new physics. When the scalar field $\phi$ acquires a nonzero $\vev$ across the bubble wall during the FOEWPT, the top-quark mass develops a spatially-varying complex phase along the bubble wall profile. This generates a $CP$-violating chiral asymmetry in the plasma in front of the wall, which biases sphaleron processes and facilitates successful baryogenesis~\cite{Cline:2012hg,Vaskonen:2016yiu,Grzadkowski:2018nbc,Ellis:2022lft}.
We will analyze the impact of this operator on the generation of baryon asymmetry in more detail in Sec.~\ref{BAU}.

\item \textbf{Departure from equilibrium:} A strongly first-order transition ensures the necessary departure from equilibrium. As bubbles of the broken phase expand and sweep through the plasma, they create a non-equilibrium environment near the bubble walls. Particles interacting with the walls can be reflected or transmitted with $CP$-violating probabilities, enabling the generation of chiral and charge asymmetries in the symmetric phase that diffuse and source baryogenesis.
\end{itemize}
Once a net baryon asymmetry is produced in the symmetric phase, it diffuses into the broken phase, where it must be protected from washout. This requires sphaleron processes to be sufficiently suppressed inside the bubbles. A commonly used criterion for this is~\cite{Quiros:1999jp,Moore:1998swa}:
\begin{equation}
    \frac{v_n}{T_n} \equiv \xi_n \gtrsim 1,
\end{equation}
where $v_n$ is the Higgs $\vev$ in the broken phase at the nucleation temperature $T_n$. This ensures that the sphaleron-induced washout is inefficient and the generated baryon asymmetry is preserved.

In summary, the dynamics of an FOEWPT provide a natural setting for realizing all the necessary ingredients of successful baryogenesis within a thermal cosmological history. While the SM falls short in both the strength of the transition and the size of $CP$ violation, various extensions, particularly those with enriched scalar sectors or additional fermions, can overcome these limitations. In this work, we will explore the viability of an SFOEWPT in our framework, mapping the regions of parameter space consistent with both baryogenesis and DM. A dedicated discussion on the realization of EWBG in our setup will be presented in Sec.~\ref{BAU}.
\subsection{Gravitational Wave Signatures from a First-Order Phase Transition}\label{GW_section}
%
In the context of this work, where the present framework potentially accommodates a strong FOPT in the early universe, it becomes highly relevant to investigate the associated stochastic GW background. Such a background can arise due to the out-of-equilibrium dynamics during the FOPT and could be detectable via cross-correlation techniques in upcoming GW interferometer experiments~\cite{Caprini:2015zlo, Cai:2017cbj, Caprini:2018mtu, Romano:2016dpx, Christensen:2018iqi}.

The production of GW during an FOPT typically involves several key mechanisms, each contributing differently to the total GW energy density spectrum, normalized by today’s critical energy density $\rho_c$ (assuming a vanishing cosmological constant). These contributions can be categorized as follows:
\begin{itemize}
    \item \textbf{Bubble wall collisions:} As expanding bubbles of the broken phase collide, the energy stored in the scalar field configurations at the walls can generate a burst of GW. However, in scenarios where the phase transition occurs within a hot plasma, the friction from the surrounding medium often prevents runaway acceleration of the walls. As a result, the GW contribution from the scalar field dynamics alone is typically subdominant and can be neglected in our analysis~\cite{Kosowsky:1991ua, Kosowsky:1992vn, Kosowsky:1992rz, Kamionkowski:1993fg, Caprini:2007xq, Huber:2008hg, Bodeker:2017cim}.

    \item \textbf{Sound waves:} The dominant source of GW in most thermal FOPT scenarios arises from acoustic waves generated by the bulk motion of the plasma. After bubble collisions, these long-lived sound waves persist and efficiently source gravitational radiation over a Hubble timescale~\cite{Hindmarsh:2013xza, Giblin:2013kea, Giblin:2014qia, Hindmarsh:2015qta}. The resulting GW signal is denoted by $\Omega_{\mathrm{sw}} h^2$, where the reduced Hubble constant is $h \equiv H_0/(100\, \mathrm{km\, s^{-1}\, Mpc^{-1}}) \approx 0.674$~\cite{DES:2017txv}.

    \item \textbf{MHD turbulence:} The violent bubble collisions and subsequent plasma motion can also trigger MHD turbulence, which acts as an additional source of GW production. Although typically subdominant compared to sound waves, the turbulence component $\Omega_{\mathrm{turb}} h^2$ can still provide a non-negligible contribution to the overall~spectrum~\cite{Caprini:2006jb, Kahniashvili:2008pf, Kahniashvili:2008pe, Kahniashvili:2009mf, Caprini:2009yp, Kisslinger:2015hua}.
\end{itemize}
Hence, the total stochastic GW signal from an FOPT can be approximately expressed as the sum of the acoustic and turbulent components:
\begin{equation}
\label{GWtotal}
\Omega_{\text{GW}}h^2 \simeq \Omega_{\text{sw}}h^2 + \Omega_{\text{turb}}h^2 \,.
\end{equation}
This cumulative signal encodes rich information about the underlying phase transition dynamics and provides a complementary probe of electroweak-scale physics beyond the SM.

To predict the resulting GW signal, one must extract five key thermodynamic and hydrodynamic parameters from the phase-transition dynamics:
nucleation temperature $T_n$, strength parameter $\alpha$, inverse duration $\beta/H_*$, relativistic degrees of freedom $g_*$, bubble wall speed $v_w$.
These inputs enable computation of the amplitude and peak frequency of the GW spectrum using semi-analytical fits calibrated from full simulations.

The nucleation temperature $T_n$ is defined via the condition:
\begin{equation}
\label{nucleation}
\int_{T_n}^\infty \frac{dT}{T} \frac{\Gamma(T)}{H(T)^4} \approx 1,
\end{equation}
indicating that one bubble nucleates per Hubble volume.
To determine the nucleation temperature $T_n$, one must evaluate the three-dimensional Euclidean action $S_3(T)$, which characterizes the tunneling probability from the false vacuum to the true vacuum~\cite{Linde:1981zj}. In our analysis, we utilize the publicly available package \texttt{CosmoTransitions}~\cite{Wainwright:2011kj} to compute the bounce solution numerically. The condition for bubble nucleation is typically expressed as the point where the probability of forming at least one bubble per Hubble volume becomes order unity. This requirement is satisfied when the nucleation rate meets the integral criterion given in Eq.~\eqref{nucleation}, and corresponds approximately to the condition 
$S_3(T_n)/T_n \approx 140$~\cite{Turner:1992tz}. Solving this equation yields the value of 
$T_n$, which represents the highest temperature at which bubble nucleation becomes efficient.

The strength parameter $\alpha$ quantifies the vacuum energy released relative to the radiation energy density:
\begin{equation}
\alpha = \frac{\epsilon}{\rho_{\rm rad}^*} = \left. \frac{1}{\rho_{\rm rad}^*} \left( - \Delta V + \frac{T}{4} \frac{d \Delta V}{dT} \right) \right|_{T_*},
\end{equation}
where 
$\Delta V = V_{\rm false} - V_{\rm true}$, and $\rho_{\rm rad}^* = \frac{\pi^2}{30} g_* T_*^4$ denotes the total radiation energy density of the plasma background. The number of relativistic degrees of freedom $g_*$ at $T=T_*$ is considered as 100.
Here, $T_*$ denotes the temperature at which the phase transition completes, corresponding approximately to $T_n$ in the absence of significant reheating.
A more accurate determination of the phase transition completion temperature can be obtained via the so-called percolation temperature. This is defined as the temperature at which a specified fraction of the universe’s volume, typically taken to be $1/e \approx 37\%$, has transitioned to the true vacuum. In cases with strong supercooling, the percolation temperature can deviate noticeably from the nucleation temperature. A full calculation of the percolation temperature is, however, beyond the scope of the present study.

The inverse time duration of the FOPT  is
\begin{equation}
\beta = - \left. \frac{d S_3}{dt} \right|_{t_*} \approx H_* T_* \left. \frac{d(S_3/T)}{dT} \right|_{T_*}.
\end{equation}
A fraction $\kappa_v$ of the released vacuum energy gets converted to bulk kinetic energy of the plasma (sound waves), and another fraction $\kappa_{\rm turb}$ to turbulence. A fitting formula for  $\kappa_v$ is~\cite{Espinosa:2010hh}:
\begin{equation}
\kappa_v(\alpha) \approx \frac{\alpha}{0.73 + 0.083 \sqrt{\alpha} + \alpha}.
\end{equation}
For the turbulence part, we need to 
 know $\kappa_{t}$, which is a fraction of $\kappa_v$.
It is expected that $\kappa_t \approx (5\sim 10) \, \kappa_v$~\cite{Hindmarsh:2015qta}, and we consider this fractional value to be 0.1 for this work.

The dominant GW contribution from the sound waves is~\cite{Hindmarsh:2013xza,Hindmarsh:2015qta,Caprini:2015zlo,Hindmarsh:2016lnk,Hindmarsh:2017gnf,Weir:2017wfa,Ellis:2018mja,Cutting:2019zws,Hindmarsh:2019phv,Ellis:2020awk,Fujikura:2021abj,Cline:2021iff,Caprini:2024hue}:
\begin{equation}
\label{soundgw}
\Omega_{\rm sw} h^2 \approx 2.65 \times 10^{-6} \, \Upsilon(\tau_{\rm sw}) \, \left( \frac{H_*}{\beta} \right) \left( \frac{\kappa_v \alpha}{1 + \alpha} \right)^2 \left( \frac{g_*}{100} \right)^{-1/3} v_w \, S_{\rm sw}(f),
\end{equation}
where the spectral shape is
\begin{equation}
S_{\rm sw}(f) = \left(\frac{f}{f_{\rm sw}}\right)^3 \left[\frac{7}{4 + 3(f/f_{\rm sw})^2}\right]^{7/2},
\end{equation}
and the peak frequency is
\begin{equation}
\label{fsw}
f_{\rm sw} \approx 1.9 \times 10^{-5} \, \mathrm{Hz} \, \frac{1}{v_w} \frac{\beta}{H_*} \left(\frac{T_n}{100\,\,\mathrm{GeV}}\right) \left(\frac{g_*}{100}\right)^{1/6}.
\end{equation}
The suppression factor from the finite lifetime of the sound waves is~\cite{Hindmarsh:2019phv}:
\begin{equation}
\Upsilon(\tau_{\rm sw}) = 1 - \frac{1}{\sqrt{1 + 2 \tau_{\rm sw} H_*}},
\end{equation}
with $\tau_{\rm sw} \approx R_*/\bar{U}_f$, $R_* \approx (8\pi)^{1/3} v_w/\beta$, and $\bar{U}_f = \sqrt{3 \kappa_v \alpha / 4}$. Here, $v_w$ denotes the velocity of the bubble wall.

The turbulence contribution is~\cite{Hindmarsh:2013xza,Hindmarsh:2015qta,Hindmarsh:2016lnk,Hindmarsh:2017gnf, Caprini:2009yp, RoperPol:2019wvy, Caprini:2015zlo, Weir:2017wfa}:
\begin{equation}
\label{GWturb}
\Omega_{\rm turb} h^2 \approx 3.35 \times 10^{-4} \left( \frac{H_*}{\beta} \right) \left( \frac{\kappa_{\rm turb} \alpha}{1 + \alpha} \right)^{3/2} \left( \frac{100}{g_*} \right)^{1/3} v_w \, S_{\rm turb}(f),
\end{equation}
with spectral shape
\begin{equation}
S_{\rm turb}(f) = \frac{(f/f_{\rm turb})^3}{[1 + (f/f_{\rm turb})]^{11/3} (1 + 8\pi f / h_*)},
\end{equation}
and peak frequency
\begin{equation}
\label{fturb}
f_{\rm turb} \approx 2.7 \times 10^{-5} \, \mathrm{Hz} \, \frac{1}{v_w} \frac{\beta}{H_*} \left(\frac{T_n}{100\,\,\mathrm{GeV}}\right)\left(\frac{g_*}{100}\right)^{1/6},
\end{equation}
where $h_* \approx 16.5 \times 10^{-6} \, \mathrm{Hz} \left(\frac{T_n}{100\,\,\mathrm{GeV}}\right)\left(\frac{g_*}{100}\right)^{1/6}$.
\section{Dark Matter Phenomenology}
\label{relicdm}
A broad experimental program seeks the particle nature of DM, including direct and indirect searches as well as collider probes at the LHC. 
In spite of intriguing anomalies, there is no definitive DM detection signal that has emerged, leading to strong bounds on its couplings to the~SM~particles.

As a reference point, consider the minimal real singlet–scalar Higgs portal with interaction
$\lambda_{hs}\,h^2 s^2$, where $s$ is a real singlet. Current direct-detection limits from LZ~\cite{LZ:2022lsv, LZCollaboration:2024lux} severely restrict this scenario: for $m_s \sim 100\,$GeV one typically requires $\lambda_{hs} \lesssim 10^{-4}$~\cite{Hooper:2025fda}. For a fixed $m_s$, a single value of $\lambda_{hs}$ reproduces the observed abundance, which then fixes the predicted rates in direct/indirect searches and at colliders. Away from the Higgs resonance, $m_s \simeq m_h/2$, where the relic density can be obtained with very small $\lambda_{hs}$, the minimal model is largely excluded by the combination of direct detection~\cite{Cline_2013,Casas:2017jjg,DiazSaez:2024nrq}, indirect detection~\cite{DeLaTorreLuque:2023fyg}, and collider constraints~\cite{Djouadi:2011aa,Arcadi:2019lka,Krnjaic:2015mbs}.
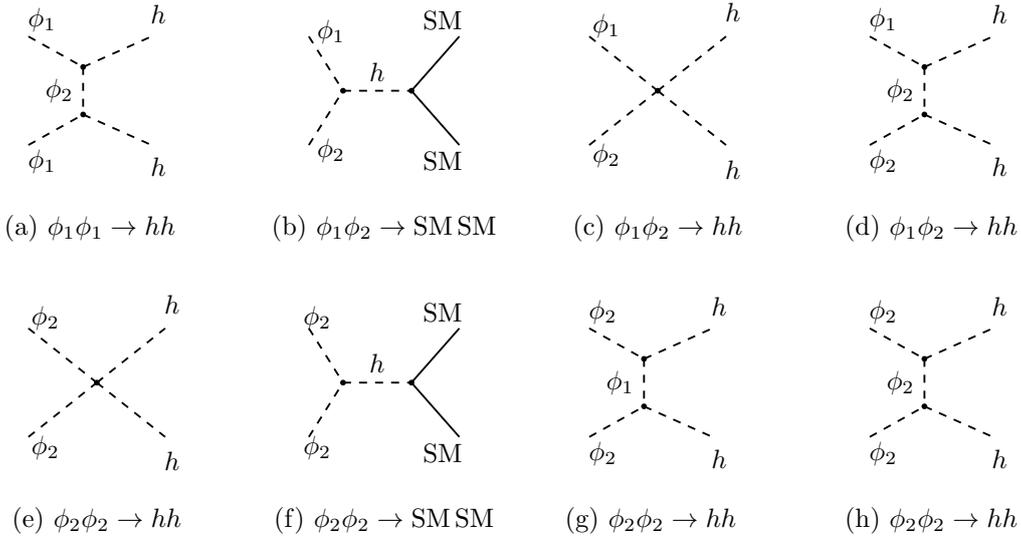
\begin{figure}[t]
\centering
\begin{tikzpicture}[line width=0.7pt, scale=0.9, >=Latex]
  \tikzset{sc/.style={dashed}, sm/.style={-}}
  \def\dx{4.1cm}      
  \def\dy{4.3cm}      
  \def\labelsep{35pt} 

  \begin{scope}[xshift=-1.5*\dx]
    \coordinate (La) at (-1,0.8);
    \coordinate (Lb) at (-1,-0.8);
    \coordinate (V1) at (-0.2,0.35);
    \coordinate (V2) at (-0.2,-0.35);
    \coordinate (Ra) at (0.8,0.8);
    \coordinate (Rb) at (0.8,-0.8);

    \draw[sc] (La) -- (V1) node[pos=0.25, above=1pt] {$\phi_1$};
    \draw[sc] (Lb) -- (V2) node[pos=0.25, below=1pt] {$\phi_1$};
    \draw[sc] (V1) -- (V2) node[midway, left] {$\phi_2$};
    \draw[sc] (V1) -- (Ra) node[pos=1.1, above=0pt] {$h$};
    \draw[sc] (V2) -- (Rb) node[pos=1.1, below=0pt] {$h$};
    \fill (V1) circle (1.2pt);\fill (V2) circle (1.2pt);

    \node at ($($(Lb)!0.5!(Rb)$)+(0,-\labelsep)$) {(a) $\phi_1\phi_1 \to hh$};
  \end{scope}

  \begin{scope}[xshift=-0.5*\dx]
    \coordinate (La) at (-1,0.8);
    \coordinate (Lb) at (-1,-0.8);
    \coordinate (VL) at (-0.5,0);
    \coordinate (VR) at (0.5,0);
    \coordinate (Ra) at (1.2,0.8);
    \coordinate (Rb) at (1.2,-0.8);

    \draw[sc] (La) -- (VL) node[pos=0.65, above=7pt] {$\phi_1$};
    \draw[sc] (Lb) -- (VL) node[pos=0.65, below=7pt] {$\phi_2$};
    \draw[sc] (VL) -- (VR) node[midway, above] {$h$};
    \draw[sm] (VR) -- node[pos=0.65, above=6pt] {$\mathrm{SM}$} (Ra);
    \draw[sm] (VR) -- node[pos=0.65, below=6pt] {$\mathrm{SM}$} (Rb);
    \fill (VL) circle (1.2pt);\fill (VR) circle (1.2pt);

    \node at ($($(Lb)!0.5!(Rb)$)+(0,-\labelsep)$) {(b) $\phi_1\phi_2 \to \mathrm{SM}\,\mathrm{SM}$};
  \end{scope}

  \begin{scope}[xshift=0.5*\dx]
    \coordinate (La) at (-1,0.8);
    \coordinate (Lb) at (-1,-0.8);
    \coordinate (V)  at (0,0);
    \coordinate (Ra) at (1,0.8);
    \coordinate (Rb) at (1,-0.8);

    \draw[sc] (La) -- (V) node[pos=0.25, above=1pt] {$\phi_1$};
    \draw[sc] (Lb) -- (V) node[pos=0.25, below=1pt] {$\phi_2$};
    \draw[sc] (V) -- (Ra) node[pos=1.1, above=0pt] {$h$};
    \draw[sc] (V) -- (Rb) node[pos=1.1, below=0pt] {$h$};
    \fill (V) circle (1.2pt);

    \node at ($($(Lb)!0.5!(Rb)$)+(0,-\labelsep)$) {(c) $\phi_1\phi_2 \to hh$};
  \end{scope}

  \begin{scope}[xshift=1.5*\dx]
    \coordinate (La) at (-1,0.8);
    \coordinate (Lb) at (-1,-0.8);
    \coordinate (V1) at (-0.2,0.35);
    \coordinate (V2) at (-0.2,-0.35);
    \coordinate (Ra) at (0.8,0.8);
    \coordinate (Rb) at (0.8,-0.8);

    \draw[sc] (La) -- (V1) node[pos=0.25, above=1pt] {$\phi_1$};
    \draw[sc] (Lb) -- (V2) node[pos=0.25, below=1pt] {$\phi_2$};
    \draw[sc] (V1) -- (V2) node[midway, left] {$\phi_2$}; 
    \draw[sc] (V1) -- (Ra) node[pos=1.1, above=0pt] {$h$};
    \draw[sc] (V2) -- (Rb) node[pos=1.1, below=0pt] {$h$};
    \fill (V1) circle (1.2pt);\fill (V2) circle (1.2pt);

    \node at ($($(Lb)!0.5!(Rb)$)+(0,-\labelsep)$) {(d) $\phi_1\phi_2 \to hh$};
  \end{scope}


  \begin{scope}[xshift=-1.5*\dx, yshift=-\dy]
    \coordinate (La) at (-1,0.8);
    \coordinate (Lb) at (-1,-0.8);
    \coordinate (V)  at (0,0);
    \coordinate (Ra) at (1,0.8);
    \coordinate (Rb) at (1,-0.8);

    \draw[sc] (La) -- (V) node[pos=0.25, above=1pt] {$\phi_2$};
    \draw[sc] (Lb) -- (V) node[pos=0.25, below=1pt] {$\phi_2$};
    \draw[sc] (V) -- (Ra) node[pos=1.1, above=0pt] {$h$};
    \draw[sc] (V) -- (Rb) node[pos=1.1, below=0pt] {$h$};
    \fill (V) circle (1.2pt);

    \node at ($($(Lb)!0.5!(Rb)$)+(0,-\labelsep)$) {(e) $\phi_2\phi_2 \to hh$};
  \end{scope}

  \begin{scope}[xshift=-0.5*\dx, yshift=-\dy]
    \coordinate (La) at (-1,0.8);
    \coordinate (Lb) at (-1,-0.8);
    \coordinate (VL) at (-0.5,0);
    \coordinate (VR) at (0.5,0);
    \coordinate (Ra) at (1.2,0.8);
    \coordinate (Rb) at (1.2,-0.8);

    \draw[sc] (La) -- (VL) node[pos=0.25, above=1pt] {$\phi_2$};
    \draw[sc] (Lb) -- (VL) node[pos=0.25, below=1pt] {$\phi_2$};
    \draw[sc] (VL) -- (VR) node[midway, above] {$h$};
    \draw[sm] (VR) -- (Ra) node[pos=0.65, above=6pt] {$\mathrm{SM}$};
    \draw[sm] (VR) -- (Rb) node[pos=0.65, below=6pt] {$\mathrm{SM}$};
    \fill (VL) circle (1.2pt);\fill (VR) circle (1.2pt);

    \node at ($($(Lb)!0.5!(Rb)$)+(0,-\labelsep)$) {(f) $\phi_2\phi_2 \to \mathrm{SM}\,\mathrm{SM}$};
  \end{scope}

  \begin{scope}[xshift=0.5*\dx, yshift=-\dy]
    \coordinate (La) at (-1,0.8);
    \coordinate (Lb) at (-1,-0.8);
    \coordinate (V1) at (-0.2,0.35);
    \coordinate (V2) at (-0.2,-0.35);
    \coordinate (Ra) at (0.8,0.8);
    \coordinate (Rb) at (0.8,-0.8);

    \draw[sc] (La) -- (V1) node[pos=0.25, above=1pt] {$\phi_2$};
    \draw[sc] (Lb) -- (V2) node[pos=0.25, below=1pt] {$\phi_2$};
    \draw[sc] (V1) -- (V2) node[midway, left] {$\phi_1$}; 
    \draw[sc] (V1) -- (Ra) node[pos=1.1, above=0pt] {$h$};
    \draw[sc] (V2) -- (Rb) node[pos=1.1, below=0pt] {$h$};
    \fill (V1) circle (1.2pt);\fill (V2) circle (1.2pt);

    \node at ($($(Lb)!0.5!(Rb)$)+(0,-\labelsep)$) {(g) $\phi_2\phi_2 \to hh$};
  \end{scope}

  \begin{scope}[xshift=1.5*\dx, yshift=-\dy]
    \coordinate (La) at (-1,0.8);
    \coordinate (Lb) at (-1,-0.8);
    \coordinate (V1) at (-0.2,0.35);
    \coordinate (V2) at (-0.2,-0.35);
    \coordinate (Ra) at (0.8,0.8);
    \coordinate (Rb) at (0.8,-0.8);

    \draw[sc] (La) -- (V1) node[pos=0.25, above=1pt] {$\phi_2$};
    \draw[sc] (Lb) -- (V2) node[pos=0.25, below=1pt] {$\phi_2$};
    \draw[sc] (V1) -- (V2) node[midway, left] {$\phi_2$}; 
    \draw[sc] (V1) -- (Ra) node[pos=1.1, above=0pt] {$h$};
    \draw[sc] (V2) -- (Rb) node[pos=1.1, below=0pt] {$h$};
    \fill (V1) circle (1.2pt);\fill (V2) circle (1.2pt);

    \node at ($($(Lb)!0.5!(Rb)$)+(0,-\labelsep)$) {(h) $\phi_2\phi_2 \to hh$};
  \end{scope}

\end{tikzpicture}

\caption{Feynman diagrams for DM annihilation channels at $f_1 \to 0$. Top row: (a) $\phi_1\phi_1\!\to hh$ (via $\phi_2$ exchange), (b) $\phi_1\phi_2\!\to \mathrm{SM}\,\mathrm{SM}$ (via $h$), (c) $\phi_1\phi_2\!\to hh$ (contact $\phi_1\phi_2 h^2$), (d) $\phi_1\phi_2\!\to hh$ (via $\phi_2$ exchange). Bottom row: (e) $\phi_2\phi_2\!\to hh$ (contact $\phi_2^2 h^2$), (f) $\phi_2\phi_2\!\to \mathrm{SM}\,\mathrm{SM}$ (via $h$), (g) $\phi_2\phi_2\!\to hh$ (via $\phi_1$ exchange), (h) $\phi_2\phi_2\!\to hh$ (via $\phi_2$ exchange). Dashed lines denote scalars ($\phi_i$, $h$); solid lines denote the SM fields. Here, the corresponding $u$-channel contributions, analogous to the $t$-channel diagrams, are not displayed.}
\label{DM-ann-Feyn-Diag}
\end{figure}

\begin{figure}[t!]
\begin{center}
    \centering
\includegraphics[width=0.55\linewidth]{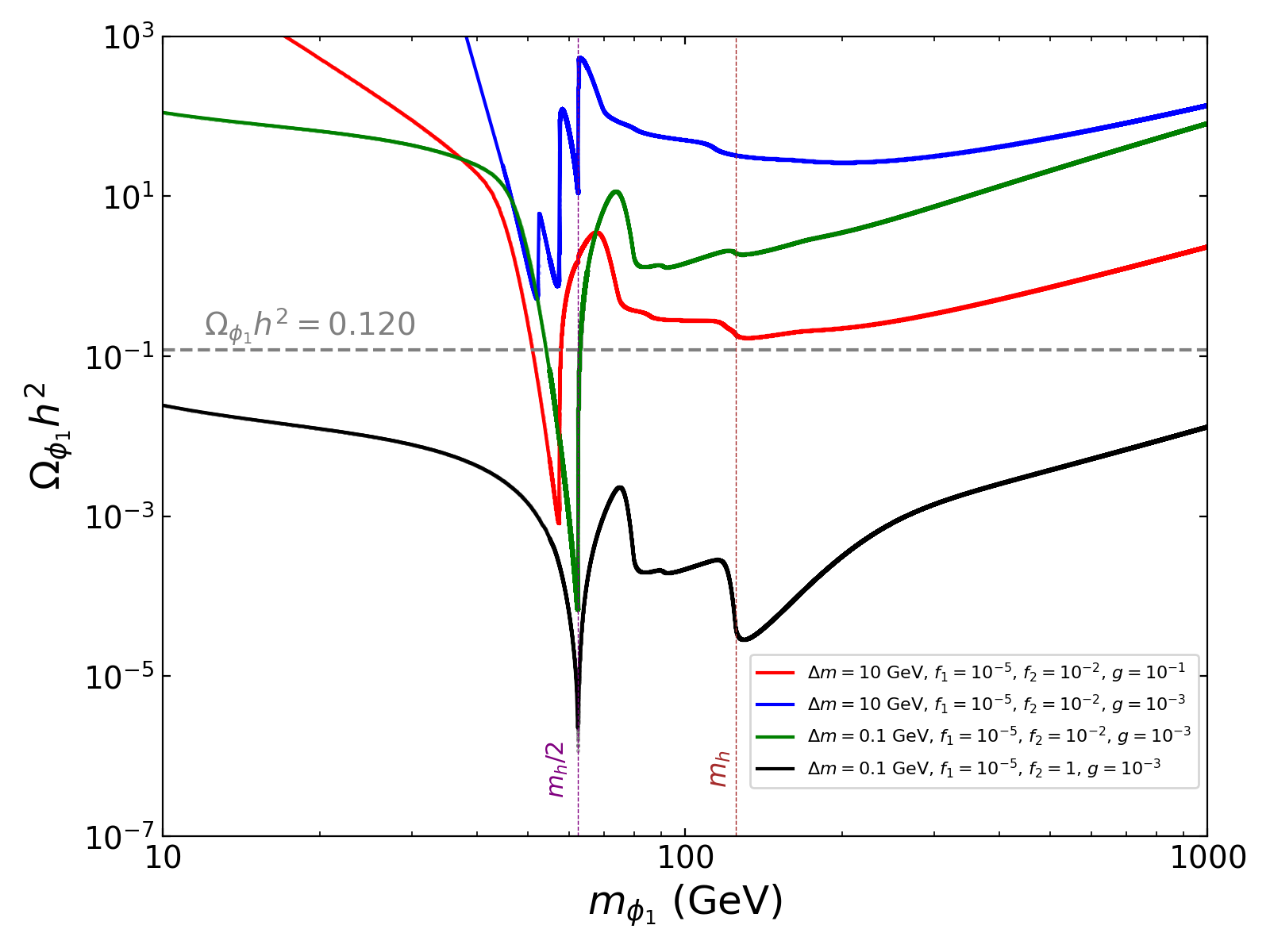}
\caption{Relic abundance $\Omega_{\phi_1} h^2$ as a function of $m_{\phi_1}$ for representative benchmark sets of $(\Delta m,\,g,\,f_2)$.
The coupling $f_1$ is fixed at $10^{-5}$, chosen small enough to ensure that elastic Higgs-mediated scattering remains well below current direct-detection limits~\cite{Hooper:2025fda,LZCollaboration:2024lux}.
The horizontal dashed line indicates the Planck measurement of the DM relic abundance~\cite{Planck:2018vyg}.
}
\label{fig:relic-plot}
\end{center}
\vspace{-0.5cm}
\end{figure}

In the setup considered in this work, after the EWSB the discrete $Z_2$ symmetry is restored, rendering the lightest singlet state, $\phi_1$, stable and a viable cold-DM candidate. To evade stringent limits from direct and indirect searches, we work in the regime of a tiny coupling $f_1$, effectively suppressing the interaction $f_1\,\phi_1^2 h^2$. The observed relic density is instead obtained through channels involving a nearby singlet $\phi_2$ via standard thermal freeze-out. In the $f_1\to 0$ limit, representative annihilation topologies are shown in Fig.~\ref{DM-ann-Feyn-Diag}. The relic abundance is primarily governed by
\begin{equation}
\{\,m_{\phi_1},\ \Delta m,\ f_2,\ g\,\}, \qquad \Delta m \equiv m_{\phi_2}-m_{\phi_1}.
\end{equation}
In this model, in the $f_1 \rightarrow 0$ limit, DM annihilates via the following channels:
\begin{enumerate}[label=(\alph*), leftmargin=*, itemsep=2pt]
\item \emph{$\phi_1$ self-annihilation:} $\phi_1\phi_1 \to hh$ via $t/u$-channel $\phi_2$ exchange; see panel~(a) of Fig.~\ref{DM-ann-Feyn-Diag}. This channel opens only for $m_{\phi_1}>m_h$ and is controlled by the inelastic coupling $g\,\phi_1\phi_2 h^2$. It has direct implications for indirect searches: present-day annihilation in the Galactic halo (or dwarf spheroidals) can yield observable gamma-ray fluxes through di-Higgs final states~\cite{Fermi-LAT:2017opo,McDaniel:2023bju,DiMauro:2022hue, Hooper:2025fda}. If the relic density is dominated by this mode, the corresponding di-Higgs rate at the Galactic Center can accommodate the observed excess of gamma-ray for $m_{\phi_1}\!\sim\!130\,$GeV~\cite{Hooper:2019xss, Hooper:2025fda}.

\item \emph{$\phi_1$–$\phi_2$ coannihilation:} For small mass splitting $\Delta m$, $\phi_2$ remains sufficiently populated to coannihilate efficiently with $\phi_1$, compensating the Boltzmann factor $e^{-\Delta m/T}$ through larger annihilation cross sections of the coannihilating species. The dominant topologies are shown in panels~(b)–(d) of Fig.~\ref{DM-ann-Feyn-Diag} and are governed by the same inelastic interaction $g\,\phi_1\phi_2 h^2$.

\item \emph{$\phi_2$ self-annihilation (assisted coannihilation):}
When $\Delta m$ is small, the heavier state can control the freeze-out of the lighter DM through a conversion–annihilation interplay. Fast inelastic scatterings with the thermal bath,
$\phi_1 + \mathrm{SM} \leftrightarrow \phi_2 + \mathrm{SM}$,
maintain chemical equilibrium between $\phi_1$ and $\phi_2$ down to temperatures near freeze-out, with a conversion rate
$\Gamma_{\rm conv}\sim n_{\rm SM}\langle\sigma' v\rangle$
enhanced by the large SM number density at $T\sim m_{\phi_1}/\mathcal{O}(20)$~\cite{Griest:1990kh}. Simultaneously, efficient $\phi_2$ self-annihilation with rate
$\Gamma_{\phi_2}\sim n_{\phi_2}\langle\sigma v\rangle$
reduces the total comoving abundance while equilibrium forces $\phi_1$ to track $\phi_2$.
The net effect, often termed assisted coannihilation~\cite{Belanger:2011ww, Dey:2016qgf}, is a substantial depletion of the eventual $\phi_1$ relic density. In the present model this dynamics is driven by the elastic coupling $f_2\,\phi_2^2 h$ and is illustrated in the lower panel~(e)-(h) of Fig.~\ref{DM-ann-Feyn-Diag}.
\end{enumerate}
To assess the relative impact of the channels, we show $\Omega_{\phi_1}h^2$ versus $m_{\phi_1}$ for benchmark choices of $(\Delta m,\,g,\,f_2)$ in Fig.~\ref{fig:relic-plot}. 
The coupling $f_1$ is fixed to a representative value $10^{-5}$, chosen sufficiently small to keep elastic Higgs-mediated scattering safely below current direct-detection bounds~\cite{Hooper:2025fda,LZCollaboration:2024lux}.
The horizontal dashed line marks the Planck value $\Omega_{\rm DM}h^2\simeq0.12$~\cite{Planck:2018vyg}; vertical dotted lines at $m_{\phi_1}\simeq m_h/2$ and $m_{\phi_1}\simeq m_h$ indicate the Higgs resonance and the $hh$ threshold. Red/blue (green/black) curves correspond to $\Delta m=10$~GeV (0.1~GeV).

A sharp suppression near $m_{\phi_1}\!\approx\!m_h/2$ arises from the $s$-channel $h$ resonance; larger $\Delta m$ shifts the dip slightly to the left of $m_h/2$, while smaller $\Delta m$ aligns it closely with $m_h/2$. Once $hh$ is kinematically open ($m_{\phi_1}\!\gtrsim\!m_h$), $\phi_2$-mediated and contact processes drive a second dip. The blue benchmark is chosen so that dips from $\phi_2\phi_2$, $\phi_1\phi_2$, and $\phi_1\phi_1$ annihilation channels are all visible.
%
%
%

The dependence on the key parameters is as follows:
\begin{itemize}
\item \textbf{$g$ (inelastic $\phi_1\phi_2 h^2$):} Larger $g$ enhances $\phi_1\phi_2$ coannihilation and $\phi_1\phi_1\!\to hh$, lowering $\Omega_{\phi_1}h^2$ and deepening the resonance dip (red vs.\ blue at $\Delta m=10$~GeV, $f_2=10^{-2}$).
\item \textbf{$\Delta m$ (splitting):} Smaller $\Delta m$ prolongs chemical equilibrium with $\phi_2$, strengthening coannihilation, shifting the dip toward $m_h/2$, and suppressing the abundance between the two markers (red/blue vs.\ green/black).
\item \textbf{$f_2$ (elastic $\phi_2^2h^2$):} For $\Delta m=0.1$~GeV and $g=10^{-3}$, increasing $f_2$ from $10^{-2}$ (green) to $1$ (black) boosts $\phi_2$-driven (assisted) coannihilation, producing the strongest depletion near $h$ and the $hh$ threshold.
\end{itemize}
At higher masses the curves rise gradually, reflecting the generic decrease of annihilation efficiency with $m_{\phi_1}$ for fixed couplings.

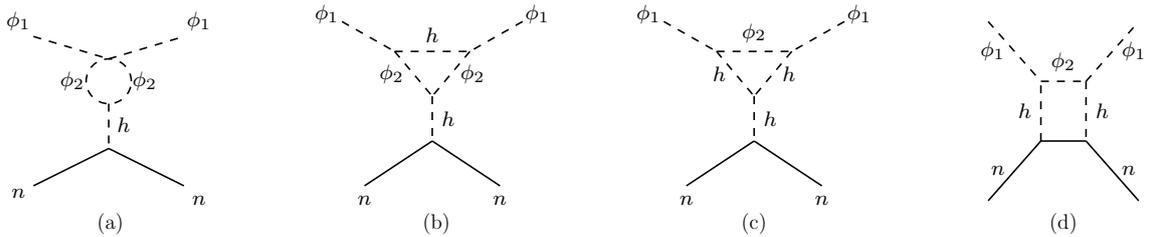
\begin{figure}[t]
\centering
\begin{tikzpicture}[line width=0.6pt, >=Latex, scale=0.55, transform shape,
                    every node/.style={font=\footnotesize},
                    baseline=(current bounding box.center)]

  \def\dx{7.0cm}
  \def\panelscale{1.8}

  \begin{scope}[xshift=-1.5*\dx, scale=\panelscale]
    \draw[dashed] (-1.8,1.0) -- (-0.8,0.7);
    \draw[solid]  (-1.8,-1.0) -- (-0.8,-0.5);

    \draw[dashed] (-0.8,0.1) -- (-0.8,-0.5);

    \draw[dashed] (-0.8,0.7) -- (0.2,1.0);
    \draw[solid]  (-0.8,-0.5) -- (0.2,-1.0);

    \draw[dashed] (-0.8,0.4) circle (0.3);
    \node at (-1.28,0.4) {\(\phi_2\)};
    \node at (-0.32,0.4) {\(\phi_2\)};

    \node at (-2.0, 1.2) {\(\phi_1\)};
    \node at (-2.0,-1.1) {\(n\)};
    \node at (-0.6,-0.18){\(h\)};
    \node at (0.4,  1.2) {\(\phi_1\)};
    \node at (0.4, -1.2) {\(n\)};
  \end{scope}
\node[font=\Large] at (-11.9,-2.7) {(a)};
  \begin{scope}[xshift=-0.6*\dx, scale=\panelscale]
    \draw[dashed] (-1.2, 1.2) -- (-0.5, 0.8); 
    \draw[dashed] ( 1.2, 1.2) -- ( 0.5, 0.8); 

    \coordinate (A) at (-0.5, 0.8);
    \coordinate (B) at ( 0.5, 0.8);
    \coordinate (C) at ( 0.0, 0.2);
    \draw[dashed] (A) -- (B) node[midway, above] {\(h\)};
    \draw[dashed] (A) -- (C) node[midway, left]  {\(\phi_2\)};
    \draw[dashed] (B) -- (C) node[midway, right] {\(\phi_2\)};

    \draw[dashed] (C) -- (0, -0.4) node[midway, right] {\(h\)};

    \draw[solid] (0, -0.4) -- (-0.9, -1.0) node[below] {\(n\)};
    \draw[solid] (0, -0.4) -- ( 0.9, -1.0) node[below] {\(n\)};

    \node at (-1.4, 1.3) {\(\phi_1\)};
    \node at ( 1.4, 1.3) {\(\phi_1\)};
  \end{scope}
\node[font=\Large] at (-4.1,-2.7) {(b)};
  \begin{scope}[xshift=0.5*\dx, scale=\panelscale]
    \draw[dashed] (-1.2, 1.2) -- (-0.5, 0.8); 
    \draw[dashed] ( 1.2, 1.2) -- ( 0.5, 0.8); 

    \coordinate (A) at (-0.5, 0.8);
    \coordinate (B) at ( 0.5, 0.8);
    \coordinate (C) at ( 0.0, 0.2);
    \draw[dashed] (A) -- (B) node[midway, above] {\(\phi_2\)};
    \draw[dashed] (A) -- (C) node[midway, left]  {\(h\)};
    \draw[dashed] (B) -- (C) node[midway, right] {\(h\)};

    \draw[dashed] (C) -- (0, -0.4) node[midway, right] {\(h\)};

    \draw[solid] (0, -0.4) -- (-0.9, -1.0) node[below] {\(n\)};
    \draw[solid] (0, -0.4) -- ( 0.9, -1.0) node[below] {\(n\)};
    \node at (-1.4, 1.3) {\(\phi_1\)};
    \node at ( 1.4, 1.3) {\(\phi_1\)};
  \end{scope}
\node[font=\Large] at (3.5,-2.7) {(c)};

  \begin{scope}[xshift=1.3*\dx, scale=\panelscale]
    \coordinate (A) at (0,  1.2);  
    \coordinate (B) at (2,  1.2);  
    \coordinate (C) at (2, -1.2);  
    \coordinate (D) at (0, -1.2);  

    \coordinate (E) at (0.7,  0.4); 
    \coordinate (F) at (1.3,  0.4); 
    \coordinate (G) at (1.3, -0.4); 
    \coordinate (H) at (0.7, -0.4); 

    \draw[dashed] (A) -- (E) node[midway,left]  {\(\phi_1\)};
    \draw[solid]  (D) -- (H) node[midway,left]  {\(n\)};
    \draw[dashed] (F) -- (B) node[midway,right] {\(\phi_1\)};
    \draw[solid]  (C) -- (G) node[midway,right] {\(n\)};

    \draw[dashed] (E) -- (F) node[midway,above] {\(\phi_2\)};
    \draw[dashed] (F) -- (G) node[midway,right] {\(h\)};
    \draw[solid]  (G) -- (H);
    \draw[dashed] (H) -- (E) node[midway,left] {\(h\)};
  \end{scope}
\node[font=\Large] at (10.9,-2.7) {(d)};
\end{tikzpicture}

    \caption{Four one-loop diagrams contributing to spin-independent scattering of \(\phi_1\) on a nucleon \(n\). Dashed lines: scalars \((\phi_{1}, \phi_2, h)\); solid lines: nucleon \(n\).}
\label{fig:DD-one-loop}
\end{figure}

Because $f_1$ can be taken to be tiny (or zero) in this model, the tree-level spin-independent scattering of $\phi_1$ on nuclei ($n$) via Higgs exchange is highly suppressed.
One might also wonder whether inelastic upscattering, 
$\phi_1 n \to \phi_2 n$, could play a role in direct detection. 
In practice, however, the mass splitting 
$\Delta m \gtrsim \mathcal{O}(100)\,\text{MeV}$ exceeds the kinetic energy 
of non-relativistic $\phi_1$, rendering such upscattering kinematically inaccessible.
The couplings that control the relic density, most notably $g$ and $f_2$, nevertheless generate one-loop contributions to direct detection. Another quartic interaction $\lambda_{12}\,\phi_1^2\phi_2^2$, which is largely irrelevant for the DM relic density estimation but can enter in loop amplitudes; in what follows we take $\lambda_{12}$ to be small so that its associated loop contribution remains subdominant. The dominant one-loop topologies are shown in Fig.~\ref{fig:DD-one-loop}: panel (a) involves $\lambda_{12}$ together with $f_2$; panel (b) depends on both $g$ and $f_2$; and panels (c)–(d) depend only on $g$. These contributions have been computed in detail in Refs.~\cite{Hooper:2025fda, DiazSaez:2024nrq}. Although they can become sizable for very large couplings, given the increasingly stringent  direct-detection upper limits from LZ, we focus on parameter regions where the loop-induced cross section lies at or below the neutrino floor.
\section{FOEWPT-favored parameter space}
\label{foptregion}
Additional scalar degrees of freedom can modify the SM Higgs potential at finite temperature so that the EWPT becomes first-order. A first-order transition requires a barrier between the symmetric (false) and broken (true) vacua. As discussed in Sec.~\ref{thermalcorr}, such a barrier can be generated by bosonic thermal loops, which induce effective cubic terms in the finite-temperature potential even when no cubic interaction is present at tree level. This motivates the BSM scenarios with extra scalars.

As a minimal baseline, consider the real singlet (\(s\)) Higgs-portal DM model with operator \(\lambda_{hs}\,h^2 s^2\). Realizing an FOEWPT in this setup typically demands a sizable portal coupling \(\lambda_{hs}\) together with a singlet mass near the electroweak scale. However, this same region is already excluded by the latest limits on the spin-independent DM–nucleon cross section from experiments such as XENONnT~\cite{XENON:2019rxp,XENON:2023cxc} and LZ~\cite{LZCollaboration:2024lux}. Consequently, a pure real-singlet Higgs-portal DM model cannot accommodate an FOEWPT while simultaneously satisfying current DM constraints.

In our setup, we extend the SM by a complex singlet, which decomposes into two real states $\phi_{1,2}$. The key advantage is that the coupling controlling tree-level direct detection can be suppressed independently of the coupling that shapes the finite-temperature Higgs potential. Concretely, taking $f_1\!\to\!0$ suppresses Higgs-mediated tree-level spin-independent scattering of the DM state $\phi_1$, while allowing the portal coupling $f_2$ (linking $\phi_2$ to $h$) to be sizable so as to strengthen the bosonic thermal effects needed for an FOEWPT.
In this construction, $f_2$ begins to contribute to direct detection at one loop, allowing comparatively large values as long as the loop-induced rate remains below current bounds.
Recent studies show that for suitable choices of $(m_{\phi_1},\,\Delta m,\,g,\,f_2)$ one can simultaneously obtain the observed DM relic density, satisfy direct-detection constraints, and realize an FOEWPT~\cite{Hooper:2025fda}. Moreover, for $m_{\phi_1}\!\sim\!130~\mathrm{GeV}$, the same region can also account for the long-standing Galactic Center gamma-ray excess via DM annihilation~\cite{Hooper:2025fda}.

\begin{figure}[t]
\centering
\includegraphics[width=0.49\linewidth]{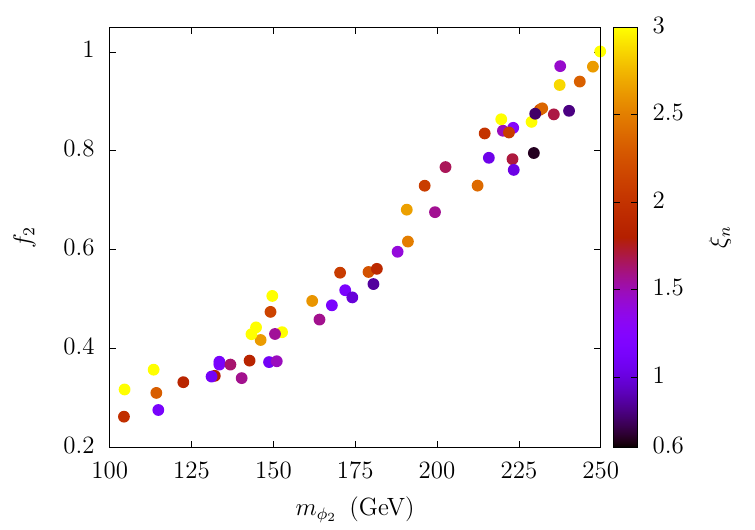}
\hfill
\includegraphics[width=0.49\linewidth]{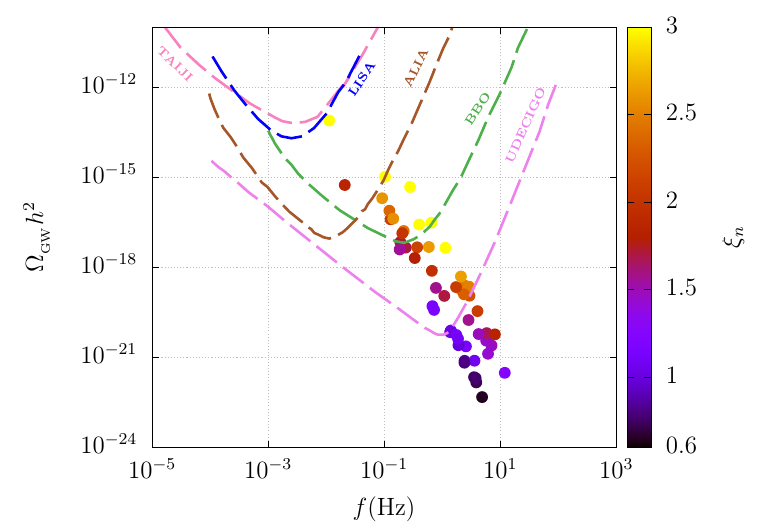}
\caption{[Left]: Scan in the $m_{\phi_2}$–$f_2$ plane showing points that realize an FOEWPT. We vary $m_{\phi_2}\in[100,250]~\mathrm{GeV}$ and $\Delta m \in[10,20]~\mathrm{GeV}$, while holding $\lambda_{1,2,12} \sim 0.5$ and $g \sim 0.05$ fixed.
[Right]: Peak amplitude $\left(\Omega_{\mathrm{GW}} h^2\right)_{\mathrm{peak}}$ versus peak frequency $f_{\mathrm{peak}}$ of the predicted stochastic GW signal from FOEWPT-allowed points. Sensitivity curves for prospective GW observatories LISA, Taiji, ALIA, BBO, and UDECIGO are overlaid for comparison.
In both panels, the color scale encodes the transition strength $\xi_n \equiv v_n/T_n$.}
\label{fig:foewpt-scan}
\end{figure}

To visualize the interplay between the portal coupling and the singlet spectrum, the left panel of Fig.~\ref{fig:foewpt-scan} shows a scan in the $(m_{\phi_2},\,f_2)$ plane restricted to points that realize an FOEWPT. We vary $m_{\phi_{1,2}}\in[100,250]~\mathrm{GeV}$ and $\Delta m\in[1,20]~\mathrm{GeV}$, while fixing $\lambda_{1,2,12} \sim 0.5$ and $g \sim 0.05$. The color scale indicates the transition strength along the Higgs direction, $\xi_n\equiv v_n/T_n$.
Using Eq.~\eqref{rho0Rsq}, in the $f_1\!\to\!0$ limit the masses satisfy
$m_{\phi_1}^2 = m_0^2 + \rho_{0_R}^2$ and
$m_{\phi_2}^2 = m_0^2 - \rho_{0_R}^2 + f_2 v_h^2$.
Thus $m_{\phi_2}^2$ increases monotonically with $f_2$, which explains the band-like correlation visible in the figure. A similar observation has been made in Refs.~\cite{Cheung:2013dca, Chala:2016ykx}. Dynamically, the allowed points tend to prefer comparatively large $f_2$, reflecting the need for sizable bosonic thermal effects to generate a barrier and achieve an FOEWPT. The scatter around the trend is driven by variations in $\Delta m$ and $m_{\phi_2}$.
While $g$ is held fixed in this plot, increasing $g$ can also influence the finite-temperature potential and, in general, shift the FOEWPT-favored band. Similarly, varying the other fixed couplings along with $g$ would select different FOEWPT-favored regions of $f_2$ for a given $m_{\phi_2}$. Note that some of the points in the plot correspond to two-step phase transitions: the first transition occurs along the singlet field directions, whereas the second transition is the FOEWPT where the Higgs develops nonzero $\vevs$ and the singlet fields return to zero $\vev$. The conditions for two-step transitions in this model are clarified below.

A distinctive feature in the potential of this model in Eq.~\eqref{Vtree} is the inelastic coupling operator
$-\frac{g}{2}\,(h^2-v_h^2)\,\phi_1\phi_2$.
What triggers a mixed singlet vacuum is the off-diagonal entry of the $(\phi_1,\phi_2)$ mass-squared matrix, which is proportional to $(h^2-v_h^2)$.
Thus, it vanishes at the EW vacuum $h=v_h$, while in the symmetric phase $h=0$ it is $-\tfrac12 g v_h^2\neq0$. From Eqs.~\eqref{mH11sq}–\eqref{mh23}, taking the \(f_1 \rightarrow 0\) limit and taking the high-temperature approximation, before EWSB the $\{\phi_1,\phi_2\}$ mass-squared matrix becomes
\begin{equation}
\label{eq:M0T}
M_\phi^2(h=0,T)=
\begin{pmatrix}
m_{\phi_1}^2+c_{22}T^2 & -\,\dfrac{g v_h^2}{2}\\[8pt]
-\,\dfrac{g v_h^2}{2} & m_{\phi_2}^2 - f_2 v_h^2 + c_{33}T^2
\end{pmatrix},
\end{equation}
where
$m_{\phi_1}^2=m_0^2+ \rho_{0_R}^2$ and $m_{\phi_2}^2=m_0^2-\rho_{0_R}^2+f_2 v_h^2$.
A vacuum with both singlet $\vevs$ nonzero exists when the origin is unstable along a mixed direction, i.e., when the lower eigenvalue of $M_\phi^2(h=0,T)$ is negative, which leads to
\begin{equation}
\label{eq:detcondT}
\big(m_{\phi_1}^2+c_{22}T^2\big)\big(m_{\phi_2}^2-f_2 v_h^2+c_{33}T^2\big)\;<\;\frac{g^2 v_h^4}{4}\,.
\end{equation}
The Daisy coefficients $c_{22}$ and $c_{33}$ are defined in Eqs.~\eqref{c22} and \eqref{c33} and depend on the model input parameters, including $f_2$, $\lambda_1$, $\lambda_2$, and $\lambda_{12}$. Taking $f_1 \rightarrow 0$ and using Eq.~\eqref{rho0Rsq}, one obtains
$ c_{22} = \tfrac{1}{24} (12 f_2 + 6 \lambda_1 + \lambda_{12})$ and
$ c_{33} = \tfrac{1}{24} (4 f_2  + 6 \lambda_2 + \lambda_{12})$.
Although Eq.~\eqref{eq:detcondT} provides the necessary condition for the singlets to develop nonzero $\vevs$ before EWSB, one can derive a useful tree-level phenomenological relation by assuming small Daisy corrections, leading to
\begin{equation}
\label{eq:detcond0}
m_{\phi_1}^2\big(m_{\phi_2}^2-f_2 v_h^2\big)\;<\;\frac{g^2 v_h^4}{4}\,.
\end{equation}
This relation highlights a distinctive correlation among $m_{\phi_1}$, $\Delta m$, $g$, and $f_2$ that selects the two-step phase-transition region. This is particularly relevant for EWBG and will be discussed in the next section.
At sufficiently high temperature, the left-hand side of \eqref{eq:detcondT} scales as $c_{22}c_{33}T^4$, so the inequality fails and all $\vevs$ vanish (symmetry restoration).

Before turning to EWBG, we briefly discuss the stochastic GW signal from such FOEWPTs, as outlined in Sec.~\ref{GW_section}.
For transitions near the electroweak scale, the redshifted spectrum today typically peaks in the mHz–Hz band, within reach of planned space-based interferometers~\cite{Grojean:2006bp,Roshan:2024qnv}.
To assess the GW signal in our setup, the right panel of Fig.~\ref{fig:foewpt-scan} shows $(\Omega_{\mathrm{GW}}h^2)_{\mathrm{peak}}$ versus $f_{\mathrm{peak}}$ for parameter points that realize an FOEWPT. Motivated by EWBG, which generally favors subsonic bubble walls to enable efficient diffusion, we fix a benchmark wall velocity $v_w=0.1$ and use this value consistently in our GW estimates (and in the EWBG analysis that follows).
For recent developments on estimating $v_w$, see Refs.~\cite{Cline:2020jre, Cline:2021iff, Laurent:2022jrs, Ekstedt:2024fyq, Li:2024mts, Carena:2025flp}.
The peak features of the GW spectrum are set primarily by the sound-wave contribution (see Eqs.~\eqref{soundgw}–\eqref{fsw}); the turbulence component (Eqs.~\eqref{GWturb}–\eqref{fturb}) provides a subleading correction and has little impact on the peak location or height in the regions we consider. The color scale encodes the transition strength $\xi_n\equiv v_n/T_n$.
A clear trend emerges: larger $\xi_n$ correlates with higher $(\Omega_{\mathrm{GW}}h^2)_{\mathrm{peak}}$ and lower $f_{\mathrm{peak}}$. In our scenario, this reflects the fact that stronger transitions nucleate at lower temperatures $T_n$, which reduce $\beta/H$, shifting the spectrum to lower frequencies (cf. Eq.~\eqref{fsw}) and enhancing the sound-wave amplitude (Eq.~\eqref{soundgw}).
From the distribution of points we infer that the predicted spectra are not accessible to LISA, whereas the bulk of the parameter space should be covered by the proposed UDECIGO experiment. Intermediate sensitivity is expected from BBO and ALIA, which could probe a subset of the viable region.

\section{Spontaneous  $CP$ violation and the generated baryon asymmetry}
\label{BAU}
Among the possible phase–transition patterns of this model, the two–step sequence is the one relevant for EWBG. At high temperature, the singlet fields first develop nonzero $\vevs$ while the Higgs remains at the origin. At lower temperature a second transition occurs: the Higgs acquires a nonzero $\vev$ and the singlet $\vevs$ return to zero, thereby restoring the discrete $Z_2$ symmetry in the electroweak–broken phase. During this second step the bubble wall interpolates between
\[
\text{ahead of the wall: }\ \langle h\rangle=0,\quad \langle \phi_{1,2}\rangle\neq0,
\qquad
\text{behind the wall: }\ \langle h\rangle\neq0,\quad \langle \phi_{1,2}\rangle=0\, .
\]
Consequently, the $CP$–violating dynamics induced by the dimension–6 operator in Eq.~\eqref{dim-6cpviolation} are localized on the wall: only within the interface do $h$ and $\phi_{1,2}$ overlap, while far from it at least one field vanishes and the source shuts off.

Once the bubbles reach a large size where curvature can be ignored and the wall expands at an approximately constant speed,
it is convenient to analyze the system in the wall rest frame using the planar-wall approximation.
Let $z$ denote the coordinate transverse to the wall. $z < 0$ ($z > 0$) denotes the electroweak symmetry broken (symmetric) phase. The background profiles are therefore one–dimensional,
$h(z), \,\, \phi_1(z), \,\, \phi_2(z)$, where $\phi=\frac{\phi_1+i \, \phi_2}{\sqrt{2}}$.
Crucially, the overlap of nonzero backgrounds occurs only within the wall region; far from the wall at least one field vanishes, so any $CP$-violating source built from their product is localized to the wall.
The static profiles can be obtain estimating the tunneling path by extremizing the following Euclidean action~\cite{Espinosa:2011ax}
\begin{equation}
\label{Euclidean}
S_E[h,\phi_1,\phi_2]=\int_{-\infty}^{+\infty}\!dz\,\bigg[\frac12\big(h'(z)\big)^2+\frac12\big(\phi_1'(z)\big)^2+\frac12\big(\phi_2'(z)\big)^2+V_{\rm eff}(h, \phi_1, \phi_2, T)\bigg],
\end{equation}
with the boundary conditions, 
\begin{align}
\label{BC}
&(h,\phi_1,\phi_2)\xrightarrow[z\to-\infty]{}\big(v_h,\,0,\,0\big),\qquad
(h,\phi_1,\phi_2)\xrightarrow[z\to+\infty]{}\big(0,\,w_{\phi}\cos\alpha,\,w_{\phi}\sin\alpha\big), \nonumber\\
&h'(\pm\infty)=\phi_1'(\pm\infty)=\phi_2'(\pm\infty)=0,
\end{align}
where $V_{\rm eff}$ is the effective finite-temperature potential defined in Eq.~\eqref{eq:Veff_full} and primes denote $\partial/\partial z$, and $ \langle h\rangle = v_h$ in the broken electroweak phase and $ \langle \phi \rangle = w_{\phi} e^{i \alpha}$ in the symmetric electroweak phase.
Thus, the wall interpolates from the electroweak-broken, singlet-symmetric phase behind the wall to the singlet-broken, Higgs-symmetric phase ahead of it, with all three fields nonzero only across the wall interface.
We estimate the  field configurations in the vicinity of the wall considering the following approximated form:
\begin{eqnarray}\label{BW}
h(z) &\equiv& \frac{v_h}{2} [1-\tanh(z/L_w)]\, ,\\ 
\phi(z) &\equiv& \frac{w_{\phi} e^{i\alpha}}{2\sqrt{2}} [1+\tanh(z/L_w)]\,,
\label{wallprfl}
\end{eqnarray}
where $L_w$ is the width of the bubble wall.
It is expected that the final path would pass or be very close to the scalar potential saddle point, which  provides the approximated expression for $L_w^2 = \frac{v_h^2 + w_{\phi}^2}{8V_b}$~\cite{Bodeker:2004ws, Espinosa:2011eu}. Here, $V_b$ denotes the barrier height of the effective potential at $T= T_n$.

In the wall rest frame, the dimension-6 interaction in Eq.~\eqref{dim-6cpviolation} makes the top mass depend on the background fields. Across the interface one has
\begin{equation}
\label{Mt}
m_t(z)\;=\;\frac{h(z)}{\sqrt{2}}\!\left[y_t+\frac{\phi(z)^2}{\Lambda^2}\right]\;\equiv\;|m_t(z)|\,e^{i\theta(z)}\, .
\end{equation}
Hence both the magnitude, $|m_t(z)|$ and the $CP$-violating phase
$\theta(z)$ vary only where \(h,\phi_1,\phi_2\) overlap--namely, inside the bubble wall.
They are given by,
\begin{eqnarray}
|m_t(z)| &=& \frac{h(z)}{\sqrt{2}}\,
\sqrt{\Big[y_t+\rho(z)\cos\!\big(2\alpha\big)\Big]^2+\Big[\rho(z)\sin\!\big(2\alpha\big)\Big]^2}\,, \\
\theta(z) &=& \tan^{-1}\!\left(\frac{\rho(z)\sin\!\big(2\alpha\big)}{\,y_t+\rho(z)\cos\!\big(2\alpha\big)}\right)\,,
\end{eqnarray}
where,
$\rho(z)\equiv \frac{|\phi(z)|^2}{2\,\Lambda^2}$ and $\phi(z)=|\phi(z)|\,e^{i\alpha}.$
Note that, far from the wall either $h=0$ (in front) or $\phi=0$ (behind), so the imaginary part vanishes.

When the top-quark mass varies across the bubble interface as $m_t(z)=|m_t(z)|e^{i\theta(z)}$, the associated phase and magnitude gradients act as $CP$-violating perturbations for tops and anti-tops traversing the wall. The resulting $CP$-odd charge densities are produced on the wall and diffuse into the electroweak-symmetric region ($z>0$), where it biases the anomalous EW sphaleron process to produce the baryon asymmetry~\cite{Cline:2012hg,Fromme:2006wx,Joyce:1994zt,Cline:1997vk,Cline:2000nw}.
In the literature, this picture is realized by solving the transport equations for chemical potentials $\mu_i$ and velocity perturbations $u_i$ where `$i$' denotes various  species of the SM~\cite{Joyce:1994zt,Cline:1997vk,Cline:2000nw}.
The most relevant SM particles in our case involve the left-handed top $t_L$, the left-handed bottom $b_L$, the right-handed top $t_R$.
The right-handed bottom can be neglected because it is generated only via a chirality flip of the left-handed bottom, which is suppressed by the bottom mass. Moreover, the Higgs perturbation typically has a subleading impact on the final asymmetry and will be omitted \cite{Fromme:2006wx}.

\begin{figure}
  \centering
  \includegraphics[width=0.49\linewidth]{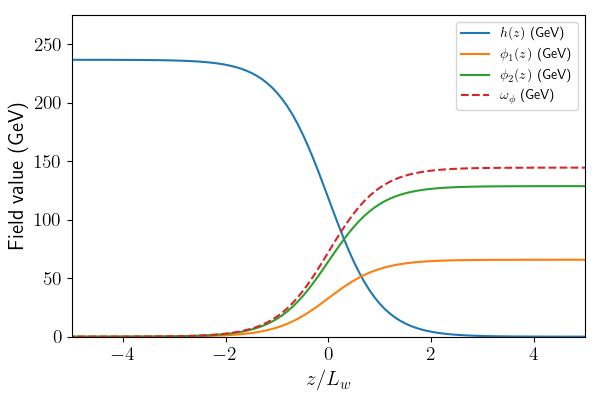}
  \hfill
  \includegraphics[width=0.49\linewidth]{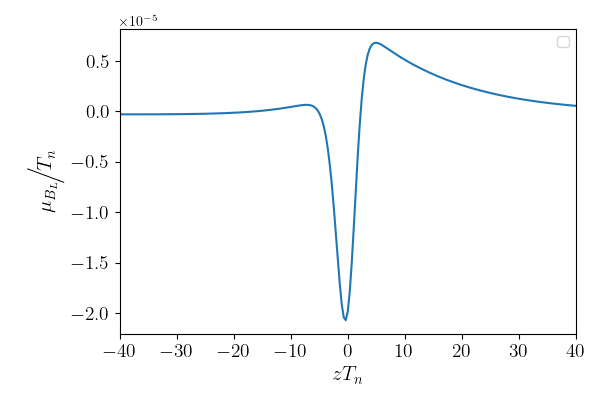}
  \caption{
  Profiles of $h(z)$, $\phi_1(z)$, and $\phi_2(z)$ across the bubble wall (left plot), and the left-handed baryon chemical potential, $\mu_{B_L}$, obtained from solving the transport equations (right plot). Both are shown as functions of the coordinate $z$, transverse to the bubble wall, at $T = T_n$ for the benchmark point given in Tab.~\ref{bptable}.}
  \label{fig:wall-mubl}
\end{figure}

In this work, we adopt the semi-classical Wenzel-Kramers-Brillouin (WKB) approach to derive the source terms and transport equations.
The top transport equations can then be written as
\cite{Fromme:2006cm,Fromme:2006wx}
\begin{subequations}
	\begin{align}
		0 =   & 3 \vw K_{1,t} \left( \partial_z \mu_{t,2} \right) + 3\vw K_{2,t} \left( \partial_z m_t^2 \right) \mu_{t,2} + 3 \left( \partial_z u_{t,2} \right) \notag
		\\ &- 3\Gamma_y \left(\mu_{t,2} + \mu_{t^c,2} + \mu_{h,2} \right) - 6\Gamma_M \left( \mu_{t,2} + \mu_{t^c,2} \right) - 3\Gamma_W \left( \mu_{t,2} - \mu_{b,2} \right) \notag
		\\ &- 3\Gamma_{ss} \left[ \left(1+9 K_{1,t} \right) \mu_{t,2} + \left(1+9 K_{1,b} \right) \mu_{b,2} + \left(1-9 K_{1,t} \right) \mu_{t^c,2} \right] \label{Eq:TransportEquations:mut} \,,\\
		0 =   & 3\vw K_{1,b} \left(\partial_z \mu_{b,2}\right) + 3 \left(\partial_z u_{b,2} \right) - 3\Gamma_y \left( \mu_{b,2} + \mu_{t^c,2} + \mu_{h,2} \right) - 3\Gamma_W \left( \mu_{b,2} - \mu_{t,2} \right) \notag \label{Eq:TransportEquations:mub}    \\
		      & - 3\Gamma_{ss} \left[ \left( 1 + 9K_{1,t}\right) \mu_{t,2} + (1+9K_{1,b}) \mu_{b,2} + (1-9K_{1,t}) \mu_{t^c,2} \right] \,,                                                                                                                      \\
		0=    & 3 \vw K_{1,t} \left( \partial_z \mu_{t^c,2} \right)  + 3\vw K_{2,t} \left( \partial_z m_t^2 \right)  \mu_{t^c,2} + 3 \left( \partial_z u_{t^c,2} \right) \notag                                                                                 \\
		      & - 3\Gamma_y \left(\mu_{t,2} + \mu_{b,2} + 2\mu_{t^c,2} + 2\mu_{h,2} \right) - 6\Gamma_M \left( \mu_{t,2} + \mu_{t^c,2} \right) \notag                                                                                                           \\
		      & - 3\Gamma_{ss} \left[ \left( 1+9 K_{1,t}\right) \mu_{t,2} + \left(1+9K_{1,b}\right) \mu_{b,2} + \left(1-9K_{1,t}\right) \mu_{t^c,2} \right] \label{Eq:TransportEquations:mutc} \,,                                                              \\
		0 =   & 4\vw K_{1,h} \left( \partial_z \mu_{h,2}\right) +
		4\left( \partial_z u_{h,2}\right) - 3\Gamma_y \left(
		\mu_{t,2} + \mu_{b,2} + 2\mu_{t^c,2} + 2\mu_{h,2} \right) -
		4\Gamma_h
		\mu_{h,2} \label{Eq:TransportEquations:muh} \,,\\
		S_t = & -3K_{4,t} \left( \partial_z \mu_{t,2}\right) + 3\vw \tilde{K}_{5,t} \left( \partial_z u_{t,2}\right) + 3\vw \tilde{K}_{6,t} \left( \partial_z m_t^2 \right) u_{t,2} + 3\Gamma_t^\mathrm{tot} u_{t,2} \label{Eq:TransportEquations:ut} \,,       \\
		0 =   & -3K_{4,b} \left( \partial_z \mu_{b,2} \right) + 3\vw \tilde{K}_{5,b} \left(\partial_z u_{b,2}\right) + 3\Gamma_b^\mathrm{tot} u_{b,2} \label{Eq:TransportEquations:ub} \,,                                                                      \\
		S_t = & -3K_{4,t} \left( \partial_z \mu_{t^c,2}\right) + 3\vw \tilde{K}_{5,t} \left( \partial u_{t^c,2}\right) + 3\vw \tilde{K}_{6,t} \left( \partial_z m_t^2\right) u_{t^c,2} + 3\Gamma_t^\mathrm{tot} u_{t^c,2} \label{Eq:TransportEquations:utc} \,, \\
		0 =   & -4K_{4,h} \left( \partial_z \mu_{h,2} \right) + 4\vw \tilde{K}_{5,h} \left( \partial_z u_{h,2} \right) + 4\Gamma_h^\mathrm{tot} u_{h,2} \label{Eq:TransportEquations:uh} \,,
	\end{align}
	\label{Eq:TransportEquations}
\end{subequations}
with the source term of the top
quark\footnote{Because of the smallness of the bottom quark mass
the source term of the bottom quark can be neglected \cite{Fromme:2006wx}.}
\begin{align}
	S_t = & -v_w K_{8,t} \partial_z \left( m_t^2 \partial_z \theta \right) + v_w K_{9,t} \left( \partial_z \theta \right) m_t^2  \left( \partial_z m_t^2\right) \label{Eq:TransportEquations:Source} \, ,
\end{align}
where various thermal averages $K_{1,i} (mi(z)/T)$ are given in Ref.~\cite{Fromme:2006wx, Cline:2011mm}.
The numerical values for the relevant reaction rates, such as, the weak/strong sphaleron~\cite{Moore:2000ara, Moore:1997im}, top Yukawa, top helicity flip, Higgs-number–violating~\cite{Huet:1995sh}, \(W\)-scattering and diffusion constants~\cite{Joyce:1994fu} at temperature \(T\) are given by,
\begin{align}
\Gamma_{\rm ws} &= 1.0\times 10^{-6}\,T, &
\Gamma_{\rm ss} &= 4.9\times 10^{-4}\,T, &
\Gamma_{y} &= 4.2\times 10^{-3}\,T, &
\Gamma_{m} &= \frac{m_t^{2}(z,T)}{63\,T}, \nonumber\\
\Gamma_{h} &= \frac{m_W^{2}(z,T)}{50\,T}, &
\Gamma_{W} &\equiv \Gamma_{h}^{\rm tot}, &
D_{q} &= \frac{6}{T}, &
D_{h} &= \frac{20}{T}.
\label{rates}
\end{align}
Solving the transport system and assuming local baryon number conservation, the effective left–handed baryon chemical potential is~\cite{Fromme:2006cm} 
\begin{align}
	\mu_{B_L} = & \frac{1}{2} \left(1+4K_{1,t}\right) \mu_{t,2} + \frac{1}{2} \left(1+4K_{1,b}\right) \mu_{b,2} - 2K_{1,t} \mu_{t^c,2} \,,
    \label{muBL}
\end{align}
which triggers the
generation of the baryon asymmetry in the
electroweak sphaleron transition. 
The baryon-to-entropy ratio generated in the unbroken phase (\(z>0\)) is then
\begin{equation}
\label{etaB}
\eta_B \equiv \frac{n_B}{s}
=\frac{405\,\Gamma_{\rm sph}}{4\pi^2 v_w g_* T}
\int_{0}^{\infty}\!dz\;\mu_{B_L}(z)\,
\exp\!\left[-\frac{45\,\Gamma_{\rm sph}}{4v_w}\,z\right],
\end{equation}
with $\Gamma_{\rm sph}\simeq 10^{-6}T$ in the electroweak-symmetric phase \cite{DOnofrio:2014rug} and $g_*=106.75$.
We take $v_w=0.1$. In the range $0.01\lesssim v_w\lesssim 0.1$, the prediction is nearly insensitive to $v_w$ as $\mu_{B_L}\propto v_w$ largely cancels the explicit $1/v_w$ in Eq.~\eqref{etaB}~\cite{Cline:2012hg}.
Later, we present a representative profile for \(\mu_{B_L}(z)\) for a benchmark point in the right plot of Fig.~\ref{fig:wall-mubl}.

The overall sign of $\eta_B$ can depend on the sign of $\sin 2\alpha$, since the $CP$–violating source term is proportional to $\sin 2\alpha$. In the $CP$-violating operator of Eq.~\eqref{dim-6cpviolation}, and in the top–mass definition of Eq.~(\ref{Mt}), one can set the Wilson coefficient `$c$' to be real and equal to 1. With this choice and for $g>0$, the singlet $\vevs$ $\phi_1$ and $\phi_2$ have the same relative sign (both positive or both negative), which corresponds to a positive $\sin 2\alpha$ and, with the mass convention of Eq.~(\ref{Mt}) and wall profiles defined in Eq.~(\ref{wallprfl}), can yield a negative $\eta_B$ in a significant range of parameter space. To obtain a relative opposite sign between  $\phi_1$ and $\phi_2$ $\vevs$, which corresponds to $\sin2\alpha < 0$, one must instead consider the region $g<0$, as shown below.

Before EWSB ($h=0$), the finite–temperature $\{\phi_1,\phi_2\}$ mass–squared matrix $M^2_{\phi}(h=0,T)$ is given in Eq.~(\ref{eq:M0T}). The mixed phase is triggered when the lower eigenvalue of \eqref{eq:M0T} becomes negative; see Eq.~\eqref{eq:detcondT}. Let $X=(x_1,x_2)^{\!\top}$ denote the (real) eigenvector associated with the lower eigenvalue $\lambda_-(T)$ of \eqref{eq:M0T}, i.e.
\begin{equation}
M^2_{\phi}(h=0,T)\,X=\lambda_-\,X\,,\qquad 
x=\begin{pmatrix}x_1\\ x_2\end{pmatrix}.
\end{equation}
From the first row,
\begin{equation}
\big(m^2_{\phi_1}+c_{22}T^2-\lambda_-\big)\,x_1-\frac{g\,v_h^2}{2}\,x_2=0
\ \ \Longrightarrow\ \
\frac{x_2}{x_1}
=\frac{m^2_{\phi_1}+c_{22}T^2-\lambda_-}{\,g\,v_h^2/2\,}.
\end{equation}
When the mixed mode goes negative we have $\lambda_-<\min\{m^2_{\phi_1}+c_{22}T^2,\ m^2_{\phi_2}-f_2 v_h^2+c_{33}T^2\}$, hence the numerator is positive. Therefore the sign of the component ratio is fixed by $g$:
\begin{equation}
\mathrm{sign}\!\left(\frac{x_2}{x_1}\right)=\mathrm{sign}(g)\,.
\end{equation}
Choosing the overall eigenvector phase so that $x_1>0$, it follows that
\begin{equation}
\mathrm{sign}(x_1 x_2)=\mathrm{sign}(g)\,.
\end{equation}
Identifying the emerging singlet $\vevs$ with the components of the light eigenvector gives the desired relation for the relative sign:
\begin{equation}
\ \mathrm{sign}(\phi_1\,\phi_2)=\mathrm{sign}(g)\ \,.
\end{equation}
This is tied to the off–diagonal entry induced by the inelastic operator, $g (h^2 -v_h^2) \phi_1 \phi_2$, which becomes $-g v_h^2 \phi_1 \phi_2$ before EWSB. To obtain opposite signs between $\phi_1$ and $\phi_2$ the off–diagonal term must be positive, which requires $g<0$. This conclusion is consistent with the analytic instability condition in Eq.~\eqref{eq:detcondT}, which selects a mixed direction precisely when the off–diagonal entry $-(g v_h^2/2)$ controls the light eigenvector of \eqref{eq:M0T}. In our parameter scan we confirm this behavior: in the two–step phase–transition scenario with $g<0$, the first transition yields nonzero $\vevs$ for $\phi_1$ and $\phi_2$ with opposite relative signs. With the top–mass convention of Eq.~\eqref{Mt}, this can lead to a positive $\eta_B$ in a large portion of parameter space.

In the $g<0$ case (with $f_1=0$), the dark–matter relic density is effectively insensitive to the sign of $g$. In our setup the inelastic coupling $g$ controls the coannihilation channels $\phi_1\phi_2 \to \text{SM\,SM},\,hh$ and enters $\phi_1\phi_1 \to hh$ via $\phi_2$ exchange, as discussed in Sec.~\ref{relicdm}. In all freeze–out amplitudes that determine $\Omega_{\phi_1}h^2$, the $g$–dependence either factors out as an overall $g$ (for $\phi_1\phi_2$ processes) or appears as $g^2$ (for $\phi_1\phi_1 \to hh$). Hence flipping $g\!\to\!-g$ multiplies the full amplitude by $-1$ or leaves it even, but in either case the squared amplitude and the thermally averaged cross sections are unchanged. The relic density therefore depends on $|g|$, $\Delta m$, and $f_2$, not on $\mathrm{sign}(g)$; the usual features (Higgs–pole dip near $m_{\phi_1}\simeq m_h/2$, additional depletion once $hh$ opens at $m_{\phi_1}\gtrsim m_h$, and the trends with $|g|$, $\Delta m$, $f_2$) are identical for $g>0$ and $g<0$.

By contrast, the one–loop spin–independent direct–detection amplitude can depend on $\mathrm{sign}(g)$ via interference among loop topologies that induce an effective $h\,\phi_1^{2}$ coupling: diagram (a) scales with $\lambda_{12}f_2$, diagram (b) with $g f_2$ (odd in $g$), and diagrams (c)–(d) with $g$ only (even in $g$ at the amplitude–squared level). Flipping $g\!\to\!-g$ reverses the linear (b) contribution relative to the even pieces, enabling destructive interference that can suppress the loop–induced spin–independent cross-section rate--while leaving freeze–out unaffected. However, in the parameter space considered here this contribution remains below current experimental sensitivity.

It is also not necessary to take $g<0$ to obtain the correct sign of $\eta_B$. Allowing a complex Wilson coefficient $c=|c|e^{i\delta}$ in Eq.~\eqref{dim-6cpviolation}~\cite{Cline:2012hg}, modifies the top mass along the wall to
\begin{equation}
m_t(z)\;=\;\frac{h(z)}{\sqrt{2}}\left[y_t+\rho(z)\,e^{i(2\alpha+\delta)}\right]\;\equiv\;|m_t(z)|\,e^{i\theta(z)}\,,
\end{equation}
with $\rho(z)\equiv |c|\,|\phi(z)|^2/(2\Lambda^2)$. The induced phase is
\begin{equation}
\theta(z)\;=\;\tan^{-1}\!\left[\frac{\rho(z)\,\sin(2\alpha+\delta)}{y_t+\rho(z)\,\cos(2\alpha+\delta)}\right]
\;\simeq\;\frac{\rho(z)}{y_t}\,\sin(2\alpha+\delta)\qquad(\rho\ll y_t)\,.
\end{equation}
Since the CP–violating source entering the transport equations scales as
$S_t\propto\partial_z\!\big(m_t^2\,\partial_z\theta\big)$, the shift $\delta\to\delta+\pi$ flips its sign and thus the sign of $\eta_B$, even for $g>0$ where $(\phi_1,\phi_2)$ acquire the same sign and $\sin2\alpha>0$. Thus, this model has the freedom to select the correct sign of $\eta_B$.

It follows from the preceding discussion that in our set-up the potential is not invariant under $\alpha\to-\alpha$. The reason is the off–diagonal inelastic operator $-\tfrac{g}{2}(h^2-v_h^2)\,\phi_1\phi_2$, which fixes the relative sign of the singlet $\vevs$ in the symmetric phase. 
Consequently, the would–be vacua at $\pm\alpha$ are not degenerate: one orientation is selected dynamically (the preferred branch is mostly set by the sign of $g$), and domain walls interpolating between $\pm\alpha$ are lifted.
This contrasts with complex–singlet extended DM models without the inelastic interaction, where the potential is symmetric under $\alpha\leftrightarrow-\alpha$ and a small explicit bias must be added to remove stable domain walls and prevent cancellation of baryon asymmetries when regions with opposite $\alpha$ collide~\cite{Grzadkowski:2018nbc}. 
In our model no such bias is required. Even though the $\pm\alpha$ degeneracy is lifted by the inelastic term, a separate class of domain wall can arise from the thermal breaking of the $Z_2$ symmetry during a two–step transition, when $\phi_{1,2}$ temporarily develop nonzero $\vevs$. Such $Z_2$ walls would eventually dominate the energy density at very late times ($T\sim10^{-7}\,$GeV)~\cite{Cline:2012hg, Espinosa:2011eu}, but in our setup they disappear earlier: the second step at the electroweak scale restores $Z_2$ and drives $\langle\phi_{1,2}\rangle\to0$, erasing these walls.
Recently, it has been shown that in two–step phase–transition scenarios, in parts of parameter space, electroweak bubbles nucleated inside the domain walls can accelerate the transition and shift $T_n$ toward higher values, closer to $T_c$. This can affect both the estimate of the BAU and the prediction of the GW spectrum~\cite{Blasi:2022woz, Agrawal:2023cgp}. A detailed analysis is left for future work; here we assume homogeneous growth of the bubble.

\begin{table}[t]
\centering
\small
\renewcommand{\arraystretch}{1.0}
\begin{tabular}{cccccccc}
\hline
$m_{\phi_1}$ [GeV] & $\Delta m$ [GeV] & $f_1$ & $f_2$ & $g$ & $\Omega h^2$  \\
\hline
133 & 17.6 & $3.4\times10^{-7}$ & 0.38 & 0.16 & 0.12  \\
\hline\hline
$\sigma_{\rm{DD}}^{\rm SI}$ (cm$^2$) &$\langle\sigma v\rangle_{\rm{ID}}^{hh}$ (cm$^3$/s) & $T_n$ (GeV) & $(h,\phi_1,\phi_2)_T$ (GeV) & $(h,\phi_1,\phi_2)_F$ (GeV) & $\eta_B$  \\
\hline
$2.26\times10^{-50}$ & $1.9\times10^{-26}$ & 95 & (237,0,0) & (0,93,182) & $8.63\times10^{-11}$  \\
\hline
\end{tabular}
\label{bptable}
\caption{Shown are the relevant model input parameters and DM observables, followed by second-step FOEWPT properties (nucleation temperature and $\vevs$ in the true (T) and false(F) minima) and the estimated baryon-to-entropy ratio \(\eta_B\).}
\end{table}

\emph{Benchmark point.-} To illustrate our scenario, we specify a benchmark (Tab.~\ref{bptable}) in which the DM mass is 
$\sim\!133~\mathrm{GeV}$ and the mass splitting is large enough that the relic density is set primarily by 
$\phi_1\phi_1\!\to hh$ via  $\phi_2$ exchange annihilation channel. 
This choice yields a present-day di-Higgs indirect-detection cross section of 
$\langle\sigma v\rangle_{\rm ID}^{hh}\simeq 1.9\times 10^{-26}~\mathrm{cm^3\,s^{-1}}$, 
consistent with interpretations of the bright, statistically significant GeV excess from the Galactic Center 
for these masses~\cite{Goodenough:2009gk,Hooper:2010mq,Hooper:2011ti,Abazajian:2012pn,Hooper:2013rwa,
Gordon:2013vta,Daylan:2014rsa,Calore:2014xka,Zhou:2014lva,Fermi-LAT:2015sau,Fermi-LAT:2017opo,
Cholis:2021rpp,DiMauro:2021raz,Hooper:2019xss,Hooper:2025fda,Hu:2025thq}. 
The elastic $\phi_1$–$h$ coupling is chosen such that the spin–independent scattering rate ($\sigma_{\rm{DD}}^{\rm SI}$) lies below 
the ``neutrino fog''~\cite{Billard:2013qya}.

This point also realizes the two-step thermal history central to our EWBG study. 
A second-order transition at $T_n=335~\mathrm{GeV}$ first breaks the $Z_2$ symmetry along the 
singlet directions so that $\langle \phi_{1,2}\rangle \neq 0$. 
At a lower temperature, $T_n=95~\mathrm{GeV}$, a first-order transition occurs in which electroweak symmetry is 
broken while the $Z_2$ is restored by driving the singlet $\vevs$ back to zero. 
The field values at the false and true minima of the second transition are listed in Tab.~\ref{bptable}. 
The corresponding bubble wall profiles $h(z)$, $\phi_{1}(z)$ and $\phi_2(z)$ are shown in the left plot of Fig.~\ref{fig:wall-mubl}. 
As discussed earlier, it can be seen that all three fields remain nonzero in certain region (within the bubble wall region), where the $CP$-violating source built.
In this benchmark scenario $L_wT_n=3.5$ and we set $v_w=0.1$. 
The generated BAU tracks the effective left-handed baryon chemical potential $\mu_{B_L}(z)$ defined in Eq.~\eqref{muBL}, 
whose profile for this point is shown in the right plot of Fig.~\ref{fig:wall-mubl}. 
To match the observed value of BAU, we set $c = 1$ and  vary the cutoff scale and fix it at $\Lambda \simeq 800~\mathrm{GeV}$ for this benchmark point.
A deep scan in the surrounding parameter space can be performed, which is expected to preserve the qualitative features of the benchmark while achieving the observed BAU with somewhat larger values of $\Lambda$.

\begin{figure}
  \centering
  \includegraphics[width=0.49\linewidth]{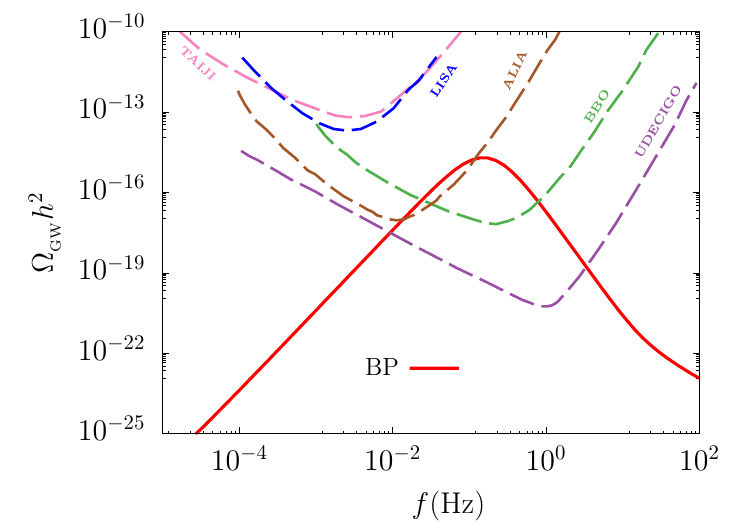}
\caption{The predicted GW energy density spectrum as a function of frequency, shown together with the experimental sensitivity curves of several detectors (LISA, Taiji, TianQin, BBO, and UDECIGO) for the benchmark point (BP) presented in Tab.~\ref{bptable}.}
  \label{fig:GWbp}
\end{figure}

Since the second phase transition is of first-order type, it also sources a stochastic GW background, which is primarily determined by four parameters: $\alpha$, $\beta/H$, $T_n$, and $v_w$, as discussed in Sec.~\ref{GW_section}.
Recent work has suggested that, for EWBG with $\beta/H \lesssim 30$, the inhomogeneities generated during the phase transition at temperatures around 100 GeV can persist until the epoch of BBN, potentially leaving observable imprints~\cite{Bagherian:2025puf}.
In our case, however, $\alpha$ and $\beta/H$ are found to be 0.06 and 845, respectively, implying that most of the inhomogeneities are expected to be erased well before BBN due to the large value of $\beta/H$.
From Fig.~\ref{fig:GWbp}, one can see that the predicted GW spectrum is relatively weak, but remains within the projected reach of BBO and UDECIGO and even grazes the sensitivity curve of ALIA.

In addition to GW signatures, collider phenomenology provides further opportunities to probe the model’s parameter space.
For $m_{\phi_{1,2}}\lesssim m_h/2$, the singlet scalars can be produced in on-shell Higgs decays, $h\to\phi_i\phi_j$; in this region the Higgs portal couplings are tightly constrained by LHC measurements of Higgs rates and by searches for invisible or exotic Higgs decays~\cite{Curtin:2013fra, Ghorbani:2014gka, Abdallah:2015ter, Abercrombie:2015wmb, Krnjaic:2015mbs, Casas:2017jjg, Boveia:2018yeb, Arcadi:2019lka, ATLAS:2025auy, Pokidova:2025jvq, Wassmer:2025kxt, Winkler:2018qyg, Guo:2025qes, Cheung:2018ave, Choi:2021nql, ATLAS:2022yvh, CMS:2023sdw, ATLAS:2023tkt, DiazSaez:2024nrq}. For $m_{\phi_{1,2}}\gtrsim m_h/2$, production proceeds mainly via an off-shell Higgs recoiling against initial-state radiation, $pp\to h^{*}j\to \phi_i\phi_j j$; with small mass splittings, the visible products from $\phi_2$ decays are often too soft to reconstruct and effectively contribute to missing transverse momentum ($\etmiss$), yielding a monojet+$\etmiss$ signature-the standard DM search strategy at the LHC~\cite{Fox:2011pm, Claude:2022rho, ATLAS:2021kxv, CMS:2021far, CMS:2022qva}.
Other processes at the LHC, such as loop-induced contributions of the singlet scalars to off-shell Higgs production, $pp \rightarrow h^{*} \rightarrow$$ZZ$, can measurably distort the differential distribution of the pair of $Z$-bosons invariant mass, providing a complementary probe of Higgs-singlet DM interaction~\cite{Goncalves:2017iub,Goncalves:2018pkt}.
Within the parameter region emphasized here, current LHC limits on this topology remain comparatively weak~\cite{Goncalves:2025snm, Guo:2025qes}. Although we do not pursue it further in this work, for very small mass splittings $\phi_2$ can be long-lived and give rise to displaced-vertex signatures~\cite{DiazSaez:2024nrq, Guo:2025qes}. 

Another important and complementary aspect of the collider phenomenology concerns direct probes of the higher-dimensional $CP$-violating operator defined in Eq.~\eqref{dim-6cpviolation}. This operator does not introduce a new light mediator or a renormalizable interaction, but instead yields a higher-dimensional correction to the top Yukawa coupling, suppressed by the scale $\Lambda^{2}$. Crucially, after the EWPT the singlet vevs vanish and both $CP$ and $Z_{2}$ symmetries are restored, so that the operator does not induce large zero-temperature effects.
Potential collider signatures associated with this operator, such as $pp \to t + \slashed{E}_T$, $t\bar t + \slashed{E}_T$, or $t\bar t H + \slashed{E}_T$, can arise depending on the event kinematics and reconstruction, and are sensitive to the dark-sector mass spectrum, the singlet mass splitting, and the validity of the EFT description at the LHC energies~\cite{Fox:2011pm, Abercrombie:2015wmb, Busoni:2013lha}.
 The existing LHC searches predominantly constrain simplified models with renormalizable couplings or light mediators, and cannot be straightforwardly mapped onto the effective interaction considered here in a model-independent manner~\cite{Abercrombie:2015wmb, Busoni:2013lha}. At present, no dedicated reinterpretation or recast of the LHC searches exists for this operator in the relevant region of parameter space, to our knowledge. Consequently, there is currently no direct collider exclusion applicable to the benchmark studied in this work, and it remains consistent with existing LHC data.
A comprehensive collider analysis of this operator--including an assessment of EFT validity, dedicated recasting of existing LHC searches, and projections for future collider sensitivities--would be highly valuable. However, such an analysis lies beyond the scope of the present work, which focuses on the DM phenomenology, EWBG, and GW signatures. We plan to pursue this direction in future work by extending our parameter scan to larger values of the cutoff scale $\Lambda$, identifying regions compatible with successful EWBG, and 
systematically exploring the collider implications of this scenario.

\section{Conclusions}
\label{Conclusions}
We presented a minimal and testable connection between DM and EWBG in a complex singlet extended Higgs-portal scenario. 
Building on earlier work that established the two-step phase transition and the direct-detection-safe, Galactic Center gamma-ray excess-compatible DM phenomenology of this model, we have now implemented EWBG within the same framework.
In this framework, the complex singlet $\phi$ splits into two non-degenerate real states $\phi_{1,2}$, both transforming identically under an imposed $Z_2$. The lighter state $\phi_1$ is stable and serves as the DM candidate. 
Importantly, the elastic $\phi_1$–$h$ coupling can be arranged to vanish or be highly suppressed, which renders the scenario safe with respect to spin-independent direct detection,
while  the heavy state \(\phi_2\) can retain a comparatively large Higgs portal coupling.
This large $\phi_2$–$h$ interaction efficiently reshapes the finite-temperature potential and favors an SFOEWPT, yet it contributes to direct detection only at loop level-remaining compatible with current bounds.
Our main findings are:
\begin{itemize}
    \item Thermal corrections can drive a two-step phase transition, which is central to the EWBG mechanism in this model. At high temperatures, the singlet acquires a nonzero $\vev$ while the Higgs remains in the symmetric phase. As the universe cools, a first-order transition into the electroweak vacuum takes place: the Higgs develops a $\vev$ while the singlet $\vev$ relaxes back to zero. This sequence yields a strongly first-order EWPT and provides the background in which the dimension-6 $CP$-violating operator localized on the bubble wall sources chiral charge for baryogenesis. Once the transition completes, both $CP$ and $Z_2$ are restored, so no residual $CP$ violation survives and EDM constraints are naturally satisfied.  We identify regions of parameter space where this pattern occurs, with the second transition being first-order and driven by a relatively large $\phi_2$-$h$ portal interaction. In combination with the wall-localized $CP$ source, this setup enables successful EWBG within the same framework that also accommodates inelastic DM, indirect detection signals, and stochastic GW.
    \item The viable EWBG regions overlap the previously established inelastic DM parameter space, where the relic density is set by coannihilation/assisted coannihilation with $\phi_2$, while remaining safe from the direct-detection bounds. For $m_{\phi_1}\gtrsim m_h$, the present-day channel $\phi_1\phi_1 \to hh$ becomes important, implying indirect-detection signatures that are compatible with the Galactic-center gamma-ray spectrum discussed in earlier work. 
    The $\phi_1$-$\phi_2$-$h$ interaction, which drives this di-Higgs production channel, also plays a key role in setting the nonzero $\vevs$ of the singlet fields in the intermediate phase of the two-step transition; this, in turn, shapes the bubble wall profiles and the magnitude of the wall-localized $CP$-violating source needed for successful EWBG.
    \item The same phase transition also generates a stochastic gravitational-wave background in the early universe. Although the spectra associated with the parameter space that produces the observed BAU via EWBG are relatively weak and typically lie below LISA’s sensitivity, they remain within reach of next-generation interferometers such as BBO and UDECIGO.
\end{itemize}
In summary, this inelastic Higgs-portal model offers a coherent framework in which (i) a two-step, strongly first-order transition with wall-localized $CP$ violation generates the baryon asymmetry, (ii) a weak-scale, direct-detection-safe DM candidate persists with predictive indirect signatures (notably $hh$ final states), and (iii) correlated GW signals provide a testable prediction for future experiments.
A representative benchmark is presented that simultaneously explains the observed Galactic Center gamma-ray excess within current astrophysical uncertainties and accounts for the observed BAU, while satisfying all existing constraints.
\section*{Acknowledgments}
The author gratefully acknowledges Dan Hooper, Gordan Krnjaic, Tong Ou and Carlos Wagner for many insightful discussions. This work is supported by the U.S. Department of Energy under contract No. DEAC02-06CH11357 at Argonne National Laboratory. The author further acknowledges the hospitality of the University of Chicago, Fermilab, the Aspen Center for Physics, and the Perimeter Institute for Theoretical Physics, where a significant part of this research was carried out. He is especially indebted to Carlos Wagner for hosting him at Perimeter Institute.

\bibliography{reference.bib}
\bibliographystyle{JHEP}

\end{document}